\documentclass[acmsmall,screen,nonacm]{acmart}
\usepackage[T1]{fontenc}
\usepackage{amsbsy,amscd,amsfonts,amssymb,amstext,amsmath,amsopn,latexsym}
\usepackage{mathtools}
\usepackage{graphicx}
\usepackage{xifthen} %conditionals for \newcommand
\usepackage{microtype}
\usepackage{listings}
\usepackage{xspace}
\usepackage{multicol}
\usepackage{fakeMnSymbol}
\usepackage{scalerel}
\usepackage{pgf,pgffor}
\usepackage{mathpartir}
\usepackage[scaled=0.81]{DejaVuSansMono}
\usepackage{tikz}
\usetikzlibrary{positioning,fit,calc}
\usepackage{tikz-cd} %for drawing commutative diagrams
\usepackage{adjustbox}
\raggedcolumns

\hypersetup{
breaklinks=false
}

\usepackage{cleveref}
\usepackage{csquotes}
\usepackage{subfig} %subfloat
\usepackage[strict]{changepage} %provides adjustwidth to avoid listings numbers protruding into margin

\lstdefinelanguage{poorman}{
  basicstyle=\selectfont\ttfamily,
  aboveskip=1pt,
  belowskip=1pt,
  literate=%
}

\lstdefinelanguage{lambdac}{
          morekeywords={let,rec,letc,ev,lete,case,in,if,from,where,then,else,await,yield,unless,on,evt,each,last,first,disjoint,match,case,when,before,after,except,cancel,aggregate,with,handler,fix,join,throw,fail,correlate,implicit,type},
          morekeywords=[2]{Top,Unit,Int,Dbl,String,Pair,Map,List,true,false},
          sensitive=true,
          mathescape=true,
          columns=fullflexible,
          keepspaces=true,
          morecomment=[l]{//},
          morecomment=[n]{/*}{*/},
          moredelim=[is][\sffamily]{!}{!},
          moredelim=[is][\rmfamily\itshape]{'}{'},
          moredelim=[is][\rmfamily\bfseries\color{ACMDarkBlue}]{?}{?},
          basicstyle=\rmfamily\mdseries,
          keywordstyle=\rmfamily\bfseries,,
          keywordstyle=[2]\sffamily,
          literate={\%p}{{'}}1 {\%q}{{?}}1 {\%x}{{!}}1 {\%bot}{{$\bot$}}1 {\%w}{{$\omega$}}1 {<}{{$\langle$}}1 {>}{{$\rangle$}}1 {\%a}{{$\alpha$}}1 {\%All}{{$\forall$}}1 {\%b}{{$\beta$}}1 {\%l}{{$\lambda$}}1 {\%L}{{$\Lambda$}}1 {\%y}{{$\gamma$}}1 {\%k}{{$\kappa$}}1 {\%v}{{$\nu$}}1 {\%e}{{$\varepsilon$}}1 {||}{{$\parallel$}}1 {\#}{{$\sharp$}}1 {=>}{{$\Rightarrow$}}1 {<:}{{$\subtpe$}}1 {<=}{{$\leq$}}1 {>=}{{$\geq$}}1 {:>}{{$\triangleright$}}1 {|->}{{$\mapsto$}}1 {[[}{{$\llbracket$}}1 {]]}{{$\rrbracket$}}1 {|+|}{{$\hcomp$}}1 {\%d}{{\$}}1
}

\lstdefinelanguage{ocaml}{
   morekeywords={assert,asr,class,closed,constraint,external,false,%
      functor,include,inherit,land,lazy,lor,lsl,lsr,lxor,method,mod,%
      module,new,open,parser,private,sig,struct,true,val,virtual,when,%
      object,ref},% TH
   morekeywords={and,as,begin,do,done,downto,else,end,exception,for,%
      fun,function,if,in,let,match,mutable,not,of,or,prefix,rec,then,%
      to,try,type,value,while,with},% NOTE: 'where' is not highlighted
   morekeywords={effect,perform,continue},
   sensitive,%
   morecomment=[n]{(*}{*)},%
   morestring=[b]",%
   basicstyle=\ttfamily,%
   keywordstyle=\ttfamily\bfseries,%
   moredelim=[is][\color{ACMRed}\underbar]{~}{~},% highlighting, or errors
   moredelim=[is][\color{ACMPurple}]{?}{?},% modules
   moredelim=[is][\color{ACMDarkBlue}]{&@}{&@},% effects
   commentstyle=\color[gray]{.4}\itshape,%
   columns=fixed
   literate=%
}

\newcommand{\commentcolor}{\color[gray]{.4}}

\ifx\TODOS\undefined
\newcommand{\todo}[1]{}
\newcommand{\seba}[1]{}
\newcommand{\diss}[1]{}
\newcommand{\note}[1]{}
\else
  \newcommand{\todo}[1]{\textcolor{red}{\textbf{#1}}}
  \newcommand{\note}[1]{\textcolor{blue}{\textbf{#1}}}
  \newcommand{\seba}[1]{\textcolor{red}{seba: #1}}
  \newcommand{\diss}[1]{}
\fi

\newcommand{\csharp}{C$\sharp$}
\newcommand{\corrl}{\textsc{Cartesius}\xspace}
\newcommand{\polyjoin}{\textsc{PolyJoin}\xspace}

\newcommand{\subtpe}{\ensuremath{\mathbin{\!\scalerel{{\lt}\!}{\colon}\!\!}}}

\newcommand*{\eg}{e.g.\@\xspace}
\newcommand*{\ie}{i.e.\@\xspace}
\newcommand*{\Ie}{I.e.\@\xspace}
\newcommand*{\Eg}{E.g.\@\xspace}

\makeatletter
\newcommand*{\etc}{%
    \@ifnextchar{.}%
        {etc}%
        {etc.\@\xspace}%
}
\makeatother

\newcommand{\syntaxcategory}[2]{{\bf #1} \hfill \ifthenelse{\isempty{#2}}{$\ $}{\fbox{$#2$}}\\[.01em]}
\newcommand{\syntaxcategorynf}[2]{{\bf #1} \hfill #2\\[.01em]}

\newcommand{\Nat}{\ensuremath{\mathbb{N}}}

\newcommand{\Set}[1]{\ensuremath{\left\{#1\right\}}} %more comfortable on my keyboard than typing \{ \}
\makeatletter
\def\slashedarrowfill@#1#2#3#4#5{%
  $\m@th\thickmuskip0mu\medmuskip\thickmuskip\thinmuskip\thickmuskip
   \relax#5#1\mkern-7mu%
   \cleaders\hbox{$#5\mkern-2mu#2\mkern-2mu$}\hfill
   \mathclap{#3}\mathclap{#2}%
   \cleaders\hbox{$#5\mkern-2mu#2\mkern-2mu$}\hfill
   \mkern-7mu#4$%
}
\def\rightslashedarrowfill@{%
  \slashedarrowfill@\relbar\relbar{\raisebox{.12em}{\tiny/}}\rightarrow}
\newcommand\xslashedrightarrow[2][]{%
  \ext@arrow 0055{\rightslashedarrowfill@}{#1}{#2}}
\def\pitcharrowfill@{%
  \arrowfill@\relbar\relbar\MNSleftpitchfork}
\newcommand\xpitchrightarrow[2][]{%
  \ext@arrow 0055{\pitcharrowfill@}{#1}{#2}}
\makeatother %TODO fork arrow still looks ugly

\newcommand{\lngl}{\ensuremath{\scalerel*{\langle}{Q}}}
\newcommand{\rngl}{\ensuremath{\scalerel*{\rangle}{Q}}}

\newcommand{\anglep}[1]{\ensuremath{\lngl #1\rngl}}

\newcommand{\thunk}[1]{\ensuremath{\{#1\}}}

\newcommand{\handler}[3]{\ensuremath{\mathbf{handler}_{#1}\ifthenelse{\isempty{#2}}{}{(#2)}\{#3\}}}

\newcommand{\eff}[1]{\ensuremath{\textcolor{ACMDarkBlue}{\mathbf{#1}}}}

\newcommand{\Type}[1]{\ensuremath{\mathsf{#1}}}

\newcommand{\ctor}[2]{\ensuremath{\mathsf{#1}\ifthenelse{\isempty{#2}}{}{\;#2}}}

\newcommand{\hastype}[2]{\ensuremath{\vdash_{\mathsf{exp}} #1: #2}}
\newcommand{\formsctx}[2]{\ensuremath{\vdash_{\mathsf{ctx}}#1: #2}}
\newcommand{\isvar}[2]{\ensuremath{\vdash_{\mathsf{var}} #1: #2}}

\newcommand{\pvars}[2]{\ensuremath{#1\leadsto #2}}
\newcommand{\formspat}[2]{\ensuremath{\vdash_{\mathsf{pat}} #1:#2}}
\newcommand{\formspatm}[3]{\ensuremath{\vdash^{#1}_{\mathsf{pat}} #2:#3}}
\newcommand{\formsext}[2]{\ensuremath{\vdash^{#1}_{\mathsf{ext}} #2}}
\newcommand{\ctxproj}[3]{\ensuremath{\vdash #1: #2\in #3}}
\newcommand{\ctxmproj}[3]{\ensuremath{\vdash #1: #2\sqsubseteq #3}}

\newlength{\reducelablen}
\newcommand{\hcomp}{\mathbin{\scalebox{0.8}{\ensuremath{\boxplus}}}}

\newcommand{\substbase}[2]{\ensuremath{#2/#1}}
\newcommand{\substloop}[1]{\ensuremath{%
\foreach \x/\v [count=\CORRLni] in {#1} {%
  \ifnum\CORRLni=1
  \substbase{\x}{\v}
  \else,\substbase{\x}{\v}\fi}
}}

\newcommand{\inflabel}[1]{\textsc{(#1)}}

\newlength{\tblsymbolwidth}
\newcommand{\tblcheck}[1]{\makebox[\tblsymbolwidth][c]{#1}}
\newcommand*\OK{\tblcheck{$\checkmark$}}
\newcommand*\SOSO{\tblcheck{$\thicksim$}}
\newcommand*\NAY{\tblcheck{\phantom{\OK}}}
\newcommand{\myparagraph}[1]{\paragraph{#1}}

\newcommand{\myhack}[1]{#1}

\makeatletter
\newbox\sf@box
\newenvironment{SubFloat}[2][]%
  {\def\sf@one{#1}%
   \def\sf@two{#2}%
   \setbox\sf@box\hbox
     \bgroup}%
  {  \egroup
    \ifx\@empty\sf@two\@empty\relax
     \def\sf@two{\@empty}
    \fi
    \ifx\@empty\sf@one\@empty\relax
      \subfloat[\sf@two]{\box\sf@box}%
    \else
      \subfloat[\sf@one][\sf@two]{\box\sf@box}%
    \fi}
\makeatother

\copyrightyear{2019}
\acmYear{2019}

\startPage{1}

\setcopyright{none}
\authorsaddresses{}

\title{Type-safe Polyvariadic Event Correlation}

 \newcommand{\TUD}{TU Darmstadt\country{Germany}}

 \author[Bra\v{c}evac]{Oliver Bra\v{c}evac}
 \affiliation{\TUD}
 \author[Salvaneschi]{Guido Salvaneschi}
 \affiliation{\TUD}
 \author[Erdweg]{Sebastian Erdweg}
 \affiliation{Johannes Gutenberg University Mainz\country{Germany}}
 \author[Mezini]{Mira Mezini}
 \affiliation{\TUD}

\newcommand{\implementationurl}{\url{http://github.com/bracevac/cartesius}}

\begin{document}

\begin{abstract}

The pivotal role that event correlation technology plays in todays applications has lead to the emergence of different families of event correlation approaches
with a multitude of specialized correlation semantics,
including computation models that support the composition and extension of different
semantics.

However, type-safe embeddings of extensible and composable event patterns
into statically-typed general-purpose programming languages have not been systematically explored so far.
This is unfortunate, as type-safe embedding of event patterns
is important to enable increased correctness of event correlation computations
as well as domain-specific optimizations.
Event correlation technology has often adopted well-known and intuitive
notations from database queries, for which approaches to type-safe embedding do
exist.
However, we argue in the
paper that these approaches, which are essentially descendants of the work on monadic
comprehensions, are not well-suited for event correlations and, thus,
cannot without further ado be reused/re-purposed
for embedding event patterns.

To close this gap we propose
\polyjoin{}, a
 novel approach to type-safe embedding for fully polyvariadic event patterns with polymorphic
  correlation semantics.
  Our approach is based on a tagless final encoding
  with uncurried higher-order abstract syntax (HOAS) representation of event patterns with
  $n$ variables, for arbitrary $n\in\Nat$.
Thus, our embedding is defined in terms of the host
language without code generation and exploits the host language type system to
model and type check the type system of the pattern language. Hence,
by construction it impossible to define ill-typed patterns.
We show that it is possible to have a purely \emph{library-level} embedding of event patterns,
in the familiar join query notation, which is not restricted to monads.  \polyjoin{} is practical,
type-safe and extensible. An implementation of it in pure multicore OCaml is readily
usable.

%%% Local Variables:
%%% mode: latex
%%% TeX-master: "report"
%%% End:

\end{abstract}

\maketitle

% !TEX root = ./report.tex

\section{Introduction}\label{polyjoins:sec:introduction}

The field of event correlation is concerned with the design,
theory, and application of pattern languages for matching events from distributed data
sources~\cite{Luckham:2001aa,Cugola:2012aa}. For example, \emph{``if within $5$
  minutes a temperature sensor reports $\geq 50^\circ$C,
  and a smoke sensor is set, trigger a fire alarm''}~\cite{Cugola:2012aa}, is a pattern that relates
sensor events by their attributes and timing.
The increasingly event-driven nature of todays applications has triggered
a lot of research and development efforts resulting in different families of event correlation approaches, \eg,
stream processing~\cite{DBLP:journals/debu/CarboneKEMHT15}, reactive
programming~\cite{Salvaneschi:2014aa}, complex event processing (CEP)~\cite{esper}.

So far, research and development efforts have been predominantly focused on
computation models that offer a wide range of specialized correlation semantics
across families as well as within the same family
(cf.~\cite{Cugola:2012aa,Bainomugisha:2013aa} for an overview).
This focus is a natural consequence of the semantic variability inherent in the domain:
Events in conjunction with time can be correlated in different ways, e.g.,
the event pattern “a followed by b” applied to the event sequence $\anglep{abb}$
may match once or twice, depending on whether the correlation computation consumes the $a$ event the first time or not. In general, it is application-specific which correlation behavior is best suited. Especially in heterogeneous computing environments, no single correlation behavior satisfies all requirements and hence different semantic variants should be expressible/composable.
To embrace this semantic variability in a principled way,
\citet{Bracevac:2018aa}
propose a computation model for programming arbitrary semantic variants of event correlation in a composable
  and extensible way. The model encodes event correlation purely in terms of algebraic effects and
  handlers~\cite{Plotkin:2003aa,Plotkin:2009aa}.

A rather neglected question concerns
the integration of event patterns into programming languages.
Such integration can be realized
by dedicated syntax and compiler extensions or by embedding, i.e., by expressing
the domain-specific language (DSL) of event patterns in terms of a host language's
linguistic concepts~\cite{Hudak:1996aa}.
Our work focuses on embeddings that  are statically type-safe and polymorphic in the sense that they allow re-targeting event pattern
syntax into different semantic representations~\cite{Hofer:2008aa}. Polymorphism
is crucial to
support the semantic diversity of event correlation.
Moreover, for practical reasons, we focus on embeddings that are implementable in
mainstream  programming languages.

Existing event correlation systems either have no language embeddings or rely on
language-integrated query techniques
 originally developed for database systems. An example in the first category is Esper~\cite{esper},
which only supports formulating event patterns/queries as strings, which are parsed and
compiled at runtime, possibly yielding runtime errors. Examples in the second category are Trill~\cite{Chandramouli:2014ug} and
Rx.Net~\cite{reactivex}, which employ LINQ~\cite{Meijer:2006aa,Cheney:2013aa} and express
event patterns in database join notation.

The reliance on database query notation for correlation patterns has historical and pragmatic
reasons: (1) some event correlation approaches have their roots in the database
community~\cite{Cugola:2012aa}, featuring similar declarative query syntax, \eg, the CQL
language~\cite{Arasu:2006aa}. (2) intuitively, one may indeed think of event patterns and join
queries as instances of a more abstract class of
operations: that which associate values originating from different sources, \eg, databases,
in-memory collections, streams, channels. (3) language-integrated query techniques
for databases are mature and readily usable for implementations of event correlation
systems.

Yet, language-integrated query techniques  for databases are not adequate for event
patterns.
Integration techniques for database languages, such as LINQ, are descendants of
monad comprehensions~\cite{Wadler:1990aa} and translate queries to instances of the monad
interface~\cite{Moggi:1991aa,Wadler:1992aa}. These techniques are statically type-checked and they are
polymorphic in the sense that they support arbitrary monad instances.
However, join queries over $n$
sources translate to $n$ variable bindings, which induce \emph{sequential} data dependencies, as this is the only interpretation
of variable bindings that the monad interface permits.
We argue that these
sequential bindings cannot express important cases of event correlation patterns, which
require a \emph{parallel} binding semantics.

Hence, we re-think how to embed event patterns into programming languages and consider
alternatives to the monad interface, in order to properly support general models for event correlation.
Specifically, adequate embeddings should support polymorphism in the semantics of event pattern
variable bindings.
To address these requirements, we propose \polyjoin{}, a novel approach for type-safe, polymorphic
embeddings of event patterns.
It embeds event patterns as a typed DSL in
OCaml and retains much of the familiar notation for database joins.

\polyjoin{} fruitfully combines two lines of research: (a) the \emph{tagless final} approach by
\citet{Carette:2009aa}, which yields type-safe, extensible and polymorphic embeddings of DSLs,
and (b) \emph{typed polyvariadic functions}, which are \emph{arity generic} (accepting a list of $n$
parameters for all $n\in\Nat$) and each of the $n$ parameters is \emph{heterogeneously-typed}~\cite{Kiselyov:2015ab}.\footnote{Polyvariadic functions were first studied by \citet{Danvy:1998aa} to encode a
statically type-safe version of the well-known \lstinline|printf| function in terms of combinators.}
Event patterns and joins naturally are instances of
polyvariadic functions: users can join $n$
heterogeneously-typed event sources, for arbitrary $n \in \Nat$.
The two lines of research complement each other well in \polyjoin{}: (1) with polyvariadic
function definitions, we can generalize Carette
et al.'s tagless embeddings from single to $n$-ary
binders in the higher-order abstract syntax (HOAS)~\cite{Huet1978,Pfenning:1988aa} encoding, and (2)
with tagless embedding, we can separate polyvariadic interfaces and polyvariadic implementations.  These
traits are crucial for event pattern embeddings, because they enable ``polymorphism in the semantics of
variable bindings''. Event patterns become DSL terms with $n$-ary
HOAS variable bindings (interface) and concrete tagless interpreters determine what the bindings
mean (implementation).

However, polyvariadic function definitions are notoriously difficult to express in
typed programming languages that are not dependently typed, like most mainstream languages.
Nevertheless, \polyjoin{} neither requires dependent types nor ad-hoc polymorphism, it requires only
forms of type constructor polymorphism and bounded polymorphism in the host language.
And it is portable: variants of the tagless final approach coincide with object
algebras~\cite{Oliveira:2012aa,Oliveira:2013aa} in OO languages.
Thus, while this paper chooses OCaml as the host language, \polyjoin{} works in principle in
other languages, both functional and OO, \eg, Haskell~\cite{Carette:2009aa}, Scala~\cite{Hofer:2008aa} or
Java~\cite{Biboudis:2015aa}.

Overall, our work shows that it is possible
to have a purely \emph{library-level} embedding of event patterns,
in the familiar join query notation, which is not restricted to monads. \polyjoin{} is
type-safe and extensible. It is readily usable as an embedding for general event correlation
systems, supporting semantically diverse event correlation. \\

\noindent
In summary, the contributions of this paper are as follows:
\begin{itemize}
\item A systematic analysis of why existing monadic style embeddings for database-style queries are
  inadequate for embedding event patterns (\Cref{polyjoins:sec:motivation}).

\item  A novel tagless final embedding for
  event patterns into mainstream programming languages, which is polyvariadic, statically type-safe,
  polymorphic, extensible, and modular (\Cref{polyjoins:sec:solution}). This embedding constitutes the core of \polyjoin{} -
  we provide a formalization and an encoding of it in pure OCaml.

\item An embedding of the \corrl{} language by \citet{Bracevac:2018aa}, yielding its first type-safe
  and fully polyvariadic implementation in multicore
  OCaml (\Cref{polyjoins:sec:case-study:-cart}).
  As already mentioned, \corrl{} supports programming arbitrary semantic variants of event
  correlation in a composable and extensible way by encoding event correlation purely in terms of
  algebraic effects and handlers, for which interesting implementation challenges arise to achieve
  polyvariadicity. Furthermore, we introduce extensions to core
  \polyjoin{} to properly support event correlation based on implicit time data associated to events.

\item An evaluation of the \polyjoin{} version of \corrl{}, comparing it against the prototype
  by \citet{Bracevac:2018aa} (\Cref{polyjoins:sec:evaluation}). Our version adds declarative pattern syntax, has exponential savings
  in code size, supports any arity and significantly reduces programmer effort when defining extensions.
  Furthermore, we discuss matters of portability and benefits of having uncurried pattern variables.
\end{itemize}

%%% Local Variables:
%%% mode: latex
%%% TeX-master: "report"
%%% End:

% !TEX root = ./report.tex

\section{Problem Statement}\label{polyjoins:sec:motivation}

In this section, we argue that
the monad interface is too limiting for the integration of event correlation systems into programming
languages.

\subsection{Event Patterns versus Join Queries}\label{polyjoins:sec:join-example}

\begin{figure}[t]
 % \centering
\begin{SubFloat}{Monadic Database Query Notation.\label{polyjoins:fig:joinpseudocode}}%
  \begin{adjustbox}{left,valign=T,margin=7pt 0pt 0pt 0pt,scale=.9}
\begin{lstlisting}[language=poorman,belowskip=-10pt,aboveskip=0pt]
from (t <- temp_sensor)
  from (s <- smoke_sensor)
    where within(s,t,5 minutes) && s && t >= 50.0
      yield (format "Fire: %f" t)
\end{lstlisting}%
  \end{adjustbox}
\end{SubFloat}\\
\begin{SubFloat}{\polyjoin{}/OCaml Version.\label{polyjoins:fig:fireexampleocaml}}%
\begin{adjustbox}{left,valign=T,margin=7pt 0pt 0pt 0pt,scale=.9}
\begin{lstlisting}[belowskip=-10pt,aboveskip=0pt]
join ((from temp_sensor) @. (from smoke_sensor) @. cnil) (*@\label{polyjoins:lst:taglessfireexample:context}@*)
       (fun ((temp,t1), ((smoke,t2), ())) -> (*@\label{polyjoins:lst:taglessfireexample:binders}@*)
          where (within t1 t2 (minutes 5.0)) %& smoke %& (temp %>= 50.0)
            (yield (format "Fire %f" temp))) (*@\label{polyjoins:lst:taglessfireexample:body_end}@*)
\end{lstlisting}
\end{adjustbox}
\end{SubFloat}%
\caption{Event Correlation Example: Fire Alarm.}
\end{figure}
%%% Local Variables:
%%% mode: latex
%%% TeX-master: "report"
%%% End:

For illustration, the pseudo-code in \Cref{polyjoins:fig:joinpseudocode} reflects how programmers could write the event pattern
at the start of \Cref{polyjoins:sec:introduction} in terms of a LINQ-like, embedded join query over
event sources.
The first two lines select/bind all events originating \lstinline[language=poorman]|from| the
temperature sensor and smoke sensor to the variables \lstinline|t| and \lstinline|s|,
respectively. The \lstinline[language=poorman]|where| clause specifies which pairs of temperature
and smoke events are relevant, \ie, those that occur at points in time at most $5$
minutes apart, the smoke event's value is \lstinline[language=poorman]|true| and the temperature
event's value is above $50^\circ$
Celsius. The \lstinline[language=poorman]|yield| clause generates a new event from each relevant
pair, in this example a string-valued event containing a warning message along with the temperature
value.  Overall, the join query correlates the \lstinline|temp_sensor| and
\lstinline|smoke_sensor| event sources and forms a new event source yielding fire alert messages.

One may think of event sources as potentially infinite sequences of discrete event notifications,
which are pairs of a value and the event's occurrence time. We write concrete event sequences within
angle brackets (\lstinline[language=poorman]|< >|). For instance,
\begin{lstlisting}[language=poorman,numbers=none]
temp_sensor: float react = < (20.0, 2), (53.5, 4), (35.0, 5), (60.2, 8) >
smoke_sensor: bool react = < (true, 9), (false, 10), (true, 12) >
\end{lstlisting}
are concrete event sequences. That is, \lstinline|temp_sensor| produces \lstinline|float|-valued events (the
temperature in degrees Celsius) and \lstinline|smoke_sensor| produces \lstinline|bool|-valued events (sensor
detects smoke or not). We name the type of event sources \lstinline|'a react|, using OCaml notation.

With the concrete event sequences above as input and assuming that the second components of the
events specify occurrence times in minutes, our example join query yields this event
sequence:\footnote{To determine the occurrence time of the events produced by a join, we merge the
  two occurrence times into the smallest interval containing both.  This merging strategy follows
  many Complex Event Processing systems~\cite{Cugola:2012aa,White:2007cr}.
  In the notation, we replace singleton intervals $[t,t]$ with just $t$.  }
\begin{lstlisting}[language=poorman,numbers=none]
< ("Fire: 53.5", [4,9]), ("Fire: 60.2", [8,9]), ("Fire: 60.2", [8,12]) >
\end{lstlisting}

\vspace{2mm}
Even though event patterns in real systems are notationally
similar to join queries in databases, there are significant semantic differences between the two:

\myparagraph{Inversion of control}
Event correlation computations are \emph{asynchronous and concurrent}.  In particular, they
  have no control when event notifications occur. Event sources run independently and produce event
  notifications at their own pace, i.e., \emph{control is inverted} as opposed to traditional
  collection or database queries, which are demand-driven. In the latter case, data is enumerated
  only if the join computation decides to access it. Dually, an event correlation computation
  passively observes and reacts to the enumeration of data.

  \myparagraph{Semantic diversity} Complex-event and stream processing systems exhibit great
  semantic diversity in event correlation behavior because events in
  conjunction with time can be correlated in different ways. For example, the event pattern
  ``\emph{$a$ followed by $b$}'' applied to the event sequence $\anglep{a\,b\,c}$
  may match once or twice, depending on whether the correlation computation consumes the $a$
  event the first time or not.  In general, it is application-specific which correlation behavior is
  best suited.  Especially in heterogeneous computing environments, no single correlation behavior
  satisfies all requirements and hence different semantic variants should be expressible/composable.

\subsection{Monadic Embeddings are Inadequate for Event Patterns}\label{polyjoins:sec:event-patterns-as}

At first sight, monads seem to be a good choice for denoting event patterns. Indeed,
monadic query embeddings have important traits, which are as important for event
patterns:

\myparagraph{(Static) type safety} They are type-safe and integrate seamlessly into the
(higher-order) host programming language. The compiler statically rejects all ill-defined queries.

\myparagraph{Polymorphic embedding} They are polymorphic embeddings~\cite{Hofer:2008aa}, because all
instances of the monad interface are admissible representations.\footnote{Strictly speaking,
  database languages require the MonadPlus interface, \ie, monads with additional zero and plus
  operations for empty bag and bag union.} Since monads encompass a large class of computations,
queries in the monadic style may denote \emph{diverse behavior}. In particular this includes
\emph{asynchronous and concurrent} computations, such as event
correlation.

A
closer inspection, though, reveals that the monadic translation of join queries cannot capture all event
correlation behaviors.

\subsubsection{Semantics of Joins in Monadic Embeddings}\label{polyjoins:sec:semant-joins-monad}
Monadic embeddings, such as LINQ, map queries to monads, \ie, type constructors $C[\cdot]$
with combinators
\begin{align*}
   \mathit{return} & :\forall \alpha. \alpha\to C[\alpha] \\
   \mathit{bind}   & :\forall \alpha.\forall\beta. C[\alpha]\to(\alpha \to C[\beta])\to C[\beta]
\end{align*}
satisfying the following laws
\begin{align}
  & \mathit{bind}\;c\;(\mathit{return}) = c    \label{polyjoins:eq:1}\\
  & \mathit{bind}\;(\mathit{return}\; x)\;f = f\; x    \label{polyjoins:eq:2}\\
  & \mathit{bind}\;(\mathit{bind}\; c\; f)\;g = \mathit{bind}\;c\;(\lambda y.\mathit{bind}\; (f\; x)\;g)    \label{polyjoins:eq:3}
\end{align}
which describe that $C[\cdot]$
models a notion of sequential computation with effects, respectively a collection type.
Accordingly,
LINQ's metatheory (cf.~\citet{Cheney:2013aa}) is specifically tailored to this
interface and its laws.  In particular, $\mathit{bind}$
models a variable binder in continuation-passing style (CPS), extracting and then binding an element of
type $\alpha$
out of the given $C[\alpha]$
shape and then continuing with the next computation step, resulting in $C[\beta]$.
Like all monadic embeddings, LINQ encodes joins by nesting invocations of
$\mathit{bind}$, so that an $n$-way
join query
\begin{lstlisting}[language=poorman,mathescape=true,numbers=none,xleftmargin=80pt]
from (x$_1$ <- r$_1$) $\cdots$ from (x$_n$ <- r$_n$) yield (x$_1,\ldots,$x$_n$)
\end{lstlisting}
denotes a nested monad computation
%\begin{center}
\[\mathit{bind}\; r_1\; (\lambda \mathit{x}_1. \mathit{bind}\; r_2\; (\lambda \mathit{x}_2. \cdots \mathit{bind}\; r_n\; (\lambda \mathit{x}_n. \; \mathit{return}\;(\mathit{x}_1, \ldots ,\mathit{x}_n))\cdots ))\]
%\end{center}
where \lstinline[language=poorman]|from|-bindings correspond to nested $\mathit{bind}$
invocations and \lstinline[language=poorman]|yield| to $\mathit{return}$.
The monad laws determine a rigid sequential selection of elements from the input sources $r_1$
to $r_n$, in the order of notation. That means, for all $i\in\Set{1,\ldots,n}$,
the monadic join computation binds an element from $r_i$
to variable $x_i$ \emph{after} it has bound all variables $x_j$, $j < i$. \\

\subsubsection{Limitations of Monadic Embeddings}\label{polyjoins:sec:limit-linqs-encod}
\begin{center}
\emph{Sequential binding is an unfaithful model of the asynchrony and
  concurrency of event sources}
\end{center}
and thus inadequate for event patterns. For example, it has
the following limitations:

\myparagraph{Unbounded increase in latency}
Suppose we embedded the fire alarm example (\Cref{polyjoins:sec:join-example}) using LINQ. The
computation is able to bind a \lstinline[language=poorman]|smoke_sensor| event to
\lstinline[language=poorman]|s| only after it has bound a \lstinline[language=poorman]|temp_sensor|
to \lstinline[language=poorman]|t|. That is, it must first unnecessarily wait on
\lstinline[language=poorman]|temp_sensor| to continue, even if a new event is already available from
\lstinline[language=poorman]|smoke_sensor|. This would increase the latency of the computation, \eg,
if past \lstinline[language=poorman]|temp_sensor| events paired with the new
\lstinline[language=poorman]|smoke_sensor| event could immediately yield a new fire alarm event. The
next \lstinline[language=poorman]|temp_sensor| event could come arbitrary late and thus delay
already available alarm events arbitrarily long.  This unbounded increase in latency is incompatible
with online data elaboration in CEP systems.
n
\myparagraph{Limited expressivity} The rigid binding order \emph{cannot} express important event
correlation behaviors which require interleaving or parallelism of bindings.
Consider correlating the sources in the fire alarm example to define a stream that \emph{always
  reflects the two most up to date values} of smoke and temperature.\footnote{This is sometimes
  referred to as \emph{Combine Latest} event correlation behavior~\cite{Bracevac:2018aa}, which
  captures the semantics of reactive programming languages.}  If one of the sources stops producing
events and the other continues, then a monadic version of this computation becomes stuck, since it
forever blocks on the non-productive source.  This example requires a ``parallel'' variable binding
semantics, where the notation order of binders bears no influence on the computation's selection
order at runtime.

\subsubsection{What Should "from" Mean?}\label{polyjoins:sec:limit-monad-embedd} In summary, monadic
query embeddings are polymorphic over the monad instance $C[\cdot]$,
where the monad interface confines to join computations with a sequential variable binding
semantics. Hence, monadic embeddings are too weak to express event patterns in join notation,
because the latter may require other semantics for variable bindings in patterns, \eg, as defined by
Applicatives~\cite{McBride:2008aa}, Arrows~\cite{Hughes:2000aa} or in the Join
Calculus~\cite{Fournet:1996aa}.  That is to say:
\begin{center}
\emph{The semantics of the \lstinline|from|-binding constitutes an additional dimension of polymorphism}.
\end{center}
Monadic embeddings fix this dimension to a single point. However, the domain of event correlation
requires embedding techniques that are parametric in this dimension as well: So that (1) systems
programmers can correctly integrate an existing event correlation engine into a programming
language, while keeping the familiar/traditional join notation. And (2), one uniform embedding
technique can accommodate diverse event correlation semantics in one application.

\subsection{Unifying Event Patterns and Join Queries}\label{polyjoins:sec:joins-type-theor}

The previous analysis suggests that we need to re-think the embedding of the join syntax into the host
programming language.
We propose that the join syntax translates to a representation which is more
general than nested monadic binds. And from now on, we let ``join'' refer to both database
joins and event correlation computations, since they both ``associate values originating
from different sources''. As a first step, we model this intuition by an informal type signature:

\begin{definition}[$n$-way join type signature]\label{polyjoins:def:joins}
At the type-level, joins are computation-transforming functions with a signature of the shape
$$S[\alpha_1]\times \cdots \times S[\alpha_n] \rightarrow S[\alpha_1\times\cdots\times \alpha_n]$$
for some type constructor $S[\cdot]$ and for all element types $\alpha_1,\ldots,\alpha_n$ and all $n\geq 0$.\myhack{\flushright\vspace{-13pt}$\blacksquare$}
\end{definition}

\noindent
That is, a join is a function merging heterogeneous $n$-tuples
of computations having a shape $S[\cdot]$
(\eg, database, event source or effect) into a computation of $n$-tuples.
For $n \in \{0,1\}$,
we obtain a constant function, respectively an identity.  The case $n > 1$
is more interesting, \eg, for $n = 2$
we obtain $$S[\alpha_1]\times S[\alpha_2]\rightarrow S[\alpha_1 \times \alpha_2],$$
which in the terminology of \citet{Mycroft:2016aa} is an \emph{effect control-flow operator}, \ie,
the (effect) $S[\cdot]$
appears left of the function arrow. This signature accommodates merge implementations that are
suitable for event correlation: They are allowed to perform the effects of the arguments in any
order, even in an interleaved or parallel fashion. For comparison, if we instantiate monadic bind
(\Cref{polyjoins:sec:semant-joins-monad}) for merging, we obtain a different control-flow
operator $$S[\alpha_1]\to (\alpha_1\to S[\alpha_1\times\alpha_2])\to S[\alpha_1\times\alpha_2].$$
By parametricity, bind continues with the next step after the input effect $S[\alpha_1]$
has been performed, with a resulting ``naked'' $\alpha_1$ which has been yielded by the shape $S[\cdot]$.
Parallel composition with other shapes is impossible.

Therefore, \Cref{polyjoins:def:joins} gives a unifying interface for both worlds: the type
signature permits both the nested monadic bind construction for database joins
(\Cref{polyjoins:sec:semant-joins-monad}), and concurrent merges, as required by event correlation.
Function values of this signature are a good target denotation for the join syntax.
Such functions are called \emph{polyvariadic}~\cite{Kiselyov:2015ab}, \ie,
functions polymorphic in both the number and (heterogeneous) type of input shapes/events sources,
reflecting that users can formulate join queries over an arbitrary, but finite number of
differently-typed sources.

We face a two-fold challenge: (1) representing polyvariadic functions as the denotation for join
syntax and (2) modeling a polymorphic embedding of join syntax that makes use of the polyvariadic
representation. Both should be expressible purely in terms of the linguistic concepts of a typed
\emph{mainstream} programming language (\eg, dependent types are forbidden).  However, such an
implementation is rewarding, because we can do it purely as a library, without being dependent on
compiler implementers to adjust their comprehension support of the language, which may never even
happen. More power to systems programmers and less burden for compiler writers!

%%% Local Variables:
%%% mode: latex
%%% TeX-master: "report"
%%% End:

% !TEX root = ./report.tex

\section{Type-safe Polyvariadic Event Patterns with \polyjoin{}}\label{polyjoins:sec:solution}

In this section, we present \polyjoin{}, an embedding of join syntax into OCaml, which is both
polymorphic and polyvariadic.  It permits more general interpretations of variable bindings in
comparison to the predominant monadic comprehensions found in modern programming languages.  For
example, \polyjoin{} admits parallel bindings that are needed for event correlation. \polyjoin{} is
lightweight: it requires neither compiler extensions, nor complicated metaprogramming, nor code
generation techniques. And it is statically type-safe: the OCaml compiler checks that joins/event
patterns cannot go wrong. Programmers can readily use \polyjoin{} to language integration and
high-level, declarative event patterns of event correlation systems.

\subsection{Fire Alarm, Revisited}\label{polyjoins:sec:fire-alarm-revisited}

As a first taste of how clients specify event patterns in \polyjoin{},
\Cref{polyjoins:fig:fireexampleocaml}
shows the \polyjoin{} version of the fire alarm example (\Cref{polyjoins:fig:joinpseudocode}),
using the \lstinline|join| form.  Note that this is \emph{pure OCaml code} and notationally
close to the original example. To avoid clashes with standard OCaml operators, we prepend
connectives in the event pattern syntax by \lstinline|%|. A
more significant difference in the notation is the separation of \lstinline|from|-bindings
(Line~\ref{polyjoins:lst:taglessfireexample:context}) from the actual body of the pattern
(Lines~\ref{polyjoins:lst:taglessfireexample:binders}-\ref{polyjoins:lst:taglessfireexample:body_end}), to
avoid nested bindings. The \lstinline|@.| symbol is a right-associative concatenation
of \lstinline|from|-bindings (cf.~\Cref{polyjoins:sec:core-polyjoin}) into a list of bindings, with
\lstinline|cnil| being the empty list of bindings.

Line~\ref{polyjoins:lst:taglessfireexample:binders} defines the pattern's variables and body in terms
of an OCaml function literal and OCaml variables, in higher-order abstract
syntax (HOAS)~\cite{Pfenning:1988aa,Huet1978}.  This way, we avoid the delicate and error-prone task of
modeling variable binding, free variables and substitution for the DSL by ourselves, instead delegating
it to the host language OCaml. We deconstruct bound events into their value and their occurrence time,
using OCaml's pattern matching, \eg, \lstinline|(temp,t1)|.

Our approach extends the work by \citet{Carette:2009aa} from single variable
HOAS bindings to uncurried $n$
variable bindings (using nested binary pairs), for arbitrary $n\in\Nat$.
We statically enforce that the number $n$
of pattern variables and their types is consistent with the number and types of supplied
\lstinline|from|-bindings: If \lstinline|temp_sensor| (resp.~\lstinline|smoke_sensor|) is an event source
of type \lstinline|float react| (resp.~\lstinline|bool react|), then pattern variable \lstinline|time|
(resp.~\lstinline|smoke|) is bound to \lstinline|float| (resp.~\lstinline|bool|) events originating from that
source in the body of the pattern.

It is impossible to define ill-typed patterns in \polyjoin{}: The type system of the host language
(OCaml) checks and enforces the correct typing of the event pattern DSL.  For example, if we added
another \lstinline|from|-binding to the pattern in \Cref{polyjoins:fig:fireexampleocaml}, then
OCaml's type checker would reject it, because bound pattern variables do not match (underlined
in the snippet below):
\begin{lstlisting}[numbers=none,belowskip=5pt,aboveskip=5pt]
join ((from temp_sensor) @. (from smoke_sensor) @. (from p_sensor) @. cnil)
      (fun ((temp,t1),((smoke,t2),~()~)) -> ...)
(* Error: This pattern matches values of type unit but a pattern was expected which matches values of type float repr * unit *)
\end{lstlisting}

\noindent In the remainder of this section, we present the core principles underlying \polyjoin{}.

\subsection{A Primer on Tagless Final Embeddings}\label{polyjoins:sec:primer-tagless-final}

\begin{figure}
\centering
\begin{minipage}[t]{.6\linewidth}%
\begin{adjustwidth}{1ex}{}%
\begin{SubFloat}{\ \ \ \ \ \label{polyjoins:fig:tagless-final-primer:symantics}}%
\begin{lstlisting}[aboveskip=0pt,belowskip=0pt]
module type ?Symantics? = sig
  type 'a repr
  val lit: int -> int repr
  val (+): int repr -> int repr -> int repr
end
\end{lstlisting}%
\end{SubFloat}%
\end{adjustwidth}%
\end{minipage}%
\begin{minipage}[t]{.4\linewidth}%
\begin{SubFloat}{\label{polyjoins:fig:tagless-final-primer:rules}}%
\begin{mathpar}
  \inferrule{n\in\mathbb{Z}}{\hastype{\mathtt{lit}\;n}{\Type{Int}}} \and  \inferrule{\hastype{e_1}{\Type{Int}}\\\hastype{e_2}{\Type{Int}}}{\hastype{e_1\;\mathtt{+}\;e_2}{\Type{Int}}}
\end{mathpar}%
\end{SubFloat}%
\end{minipage}\\
\begin{minipage}[t]{.6\linewidth}%
\begin{adjustwidth}{1ex}{}%
\begin{SubFloat}{\label{polyjoins:fig:tagless-final-primer:num}}%
\begin{lstlisting}[aboveskip=0pt,belowskip=0pt]
module ?Num? = struct
  type 'a repr = 'a
  let lit n = n
  let (+) x y = plus x y

end
\end{lstlisting}%
\end{SubFloat}%
\end{adjustwidth}%
\end{minipage}%
\begin{minipage}[t]{.4\linewidth}%
\begin{adjustwidth}{1ex}{}%
\begin{SubFloat}{\label{polyjoins:fig:tagless-final-primer:pp}}%
\begin{lstlisting}[aboveskip=0pt,belowskip=0pt]
module ?PP? = struct
  type 'a repr = string
  let lit n = sprintf "<%d>" n
  let (+) x y =
    sprintf "(%s + %s)" x y
end
\end{lstlisting}%
\end{SubFloat}%
\end{adjustwidth}%
\end{minipage}%
% \\
% \begin{minipage}[t]{.6\linewidth}%
% \begin{lstlisting}[aboveskip=0pt,belowskip=0pt]
% module type Symantics' = sig
%   include Symantics
%   val b_lit: bool -> bool repr
%   val (&): bool repr -> bool repr -> bool repr
% end
% \end{lstlisting}%
% \subcaption{}
% \end{minipage}%
% \begin{minipage}[t]{.4\linewidth}%
% \begin{lstlisting}[aboveskip=0pt,belowskip=0pt]
% module NumBool = struct
%   include Num
%   let b_lit b = b
%   let (&) x y = x && y
% end
% \end{lstlisting}%
% \subcaption{}%
% \end{minipage}%
\myhack{\vspace{-10pt}}\caption{Basic Tagless Final Examples.}\label{polyjoins:fig:tagless-final-primer}\myhack{\vspace{-10pt}}
\end{figure}

%%% Local Variables:
%%% mode: latex
%%% TeX-master: "report"
%%% End:

We briefly recapitulate the tagless final approach by \citet{Carette:2009aa} in the following.
Readers familiar with the topic may skip this section.

Traditionally, interpreters for DSLs are defined by structurally recursive functions (initial
algebras) translating abstract syntax terms into a semantic domain, \ie, the interpreter function folds the
abstract syntax terms. The latter are modeled by data types, ``tagging'' nodes with data
constructors, and deconstructed by the interpreter function via pattern matching.  In contrast,
tagless final (1) encodes DSL terms with (typed) functions instead of data, in the style of
\citet{Reynolds:1978aa} and (2) decouples interface and implementation (the interpreter) of these
functions, obtaining a term representation which is indexed by their denotation.

For example, the OCaml module signature in \Cref{polyjoins:fig:tagless-final-primer:symantics}
defines a tagless DSL for arithmetic expressions. This language supports integer constants and
addition, via the syntax functions \lstinline|lit| and addition \lstinline|(+)|, in infix
notation. We abstract over the concrete semantic representation of arithmetic expressions with the
type constructor \lstinline|'a repr| (read: expression of DSL type \lstinline|'a|). The type
signatures of these syntax functions embed both the grammar and the typing rules of the DSL in the
host language's type system, using phantom types~\cite{Leijen:1999aa}.  These signatures
straightforwardly correspond to a bottom-up
reading of natural deduction rules, with return type being conclusions and arguments being premises
(\Cref{polyjoins:fig:tagless-final-primer:rules}). That is, tagless final models
\emph{intrinsically-typed} DSLs, where it is impossible to define ill-typed terms, by construction.
We write $\hastype{M}{A}$ for expression typing, assigning DSL expression $M$ the DSL type $A$.

In OCaml, we represent (groups of) concrete DSL terms as functors,
accepting a \lstinline|Symantics| module:
\begin{lstlisting}[aboveskip=0pt,belowskip=0pt]
module ?Exp?(?S?: ?Symantics?) = struct
  open ?S?
  let exp1 = (lit 1) + (lit 2)
  let exp2 x y = exp1 + (lit 3) + x + y
end
\end{lstlisting}
In this example, \lstinline|exp1| is a closed DSL term having host language type
\lstinline|int S.repr| and \lstinline|exp2| a DSL term with two free variables, having the type
\begin{lstlisting}[aboveskip=0pt,belowskip=0pt,numbers=none,xleftmargin=100pt]
int ?S?.repr -> int ?S?.repr -> int ?S?.repr
\end{lstlisting}

The interpretation of DSL terms
depends on modules conforming to the \lstinline|?Symantics?| signature above, \ie, DSL terms are
parametric in their denotation/interpreter \lstinline|?S?|.
For brevity, we will not explicitly show the surrounding functor boilerplate in subsequent example terms.
\Cref{polyjoins:fig:tagless-final-primer:num} and \Cref{polyjoins:fig:tagless-final-primer:pp} show
two example interpreters/denotations for our arithmetic expressions DSL. \lstinline|Num|
interprets arithmetic expressions as OCaml's built-in integers and functions,
whereas
\lstinline|?PP?| interprets them as their string representation:
\begin{lstlisting}
module ?En? = ?Exp?(?Num?);; (* Instantiate Num interpretation *)
module ?Es? = ?Exp?(?PP?);;  (* Instantiate PP interpretation *)
?En?.exp1;; (* yields int Num.repr = 3 *)
?Es?.exp1;; (* yields int PP.repr = "(<1> + <2>)" *)
\end{lstlisting}

Furthermore, the tagless final approach supports modular extensibility of DSLs.  For example, the
DSL for arithmetic expressions can be extended to further include boolean expressions and
connectives:
\begin{lstlisting}
module type ?NumBoolSym? = sig
  include ?Symantics?
  val b_lit: bool -> bool repr
  val (&): bool repr -> bool repr -> bool repr
end
\end{lstlisting}
In a similar fashion, these additional functions can be separately implemented in a module and
combined with the existing interpreters, using module composition, to yield extensions of the
interpreters, such as \lstinline|?Num?| and \lstinline|?PP?|.  We leave their extension with booleans as
an exercise to the reader.

Finally (no pun intended), while the focus of this paper is polyvariadic and polymorphic event
pattern embeddings, basing \polyjoin{} on the tagless final approach enables promising future
applications for our line of research: in conjunction with multi-stage programming, the approach in
principle supports modular, library-level compilation pipelines for DSLs, \eg, as exemplified in
\cite{Carette:2009aa} and \cite{Suzuki:2016aa}.

\subsection{Core \polyjoin{}}\label{polyjoins:sec:core-polyjoin}

\begin{figure}
\syntaxcategorynf{Expressions and Patterns}{\fbox{$\hastype{M}{A}$}  \fbox{$\formspat{P}{A}$}}
\begin{mathpar}
\inferrule[where]{\hastype{M}{\Type{Bool}}\\ \formspat{P}{A}}{\formspat{\mathtt{where}\;M\;P}{A}}
\and
\inferrule[yield]{\hastype{M}{A}}{\formspat{\mathtt{yield}\;M}{A}}
\and
\inferrule[join]{\formsctx{\Pi}{\vec{A}}\\ \pvars{\vec{A}}{\vec{B}}\\\inferrule*{\inferrule*[vdots=1em]{}{[\overrightarrow{\hastype{x}{B}}]}}{\formspat{P}{C}}}{\hastype{\mathtt{join}\;\Pi\;(\vec{x}.P)}{\Type{Shape}[C]}}
\end{mathpar}
\syntaxcategorynf{Context Formation}{\fbox{$\isvar{V}{A}$} \fbox{$\formsctx{\Pi}{\vec{A}}$}}
\begin{mathpar}
  \inferrule*[Right=from,vcenter]{\hastype{M}{\Type{Shape}[A]}}{\isvar{\mathtt{from}\; M}{A}}
  \and
  \inferrule*[Right=cnil,vcenter]{}{\formsctx{\mathtt{cnil}}{\Type{\varnothing}}}
  \and
  \inferrule*[Right=cat,vcenter]{\isvar{V}{B}\\ \formsctx{\Pi}{\vec{A}}}{\formsctx{V \mathbin{\mathtt{@.}}\Pi}{B,\vec{A}}}
\end{mathpar}
\syntaxcategory{Shape Translation}{$\pvars{\vec{A}}{\vec{B}}$}\myhack{\vspace{-3ex}}
$\inferrule*[Right=id]{}{\pvars{\vec{A}}{\vec{A}}}$%
\myhack{\vspace{-10pt}}\caption{Core Syntax and Typing Rules of \polyjoin{}.}\label{polyjoins:fig:polyjoin-core}
\end{figure}
%%% Local Variables:
%%% mode: latex
%%% TeX-master: "report"
%%% End:

\begin{figure}
%\centering
\begin{adjustbox}{left,margin=0pt 0pt 15pt 0pt,valign=T,scale=.9}
\begin{lstlisting}[columns=fixed,aboveskip=0pt,belowskip=0pt]
module type ?Symantics? = sig
  type 'a shape (* $\commentcolor\Type{Shape}[\cdot]$ Constructor *)
  (* Judgments $\textit{\commentcolor (cf.~\Cref{polyjoins:fig:polyjoin-core}})$: *)
  type 'a repr     (* $\commentcolor\hastype{\cdot}{A}$ *)
  type 'a pat      (* $\commentcolor\formspat{\cdot}{A}$ *)
  type ('a,'b) ctx (* combination of $\commentcolor\formsctx{\cdot}{A}\ \textit{and}\ \pvars{A}{B}$ *)
  type 'a var      (* $\commentcolor\isvar{\cdot}{A}$ *)
  (* Context Formation and Shape Translation: *)
  val from: 'a shape repr -> 'a var
  val cnil: (unit,unit) ctx
  val (@.): 'a var -> ('c, 'd) ctx -> ('a * 'c, 'a repr * 'd) ctx
  (* Expressions and Patterns: *)
  val yield: 'a repr -> 'a pat
  val where: bool repr -> 'a pat -> 'a pat
  val join: ('a, 'b) ctx -> ('b -> 'c pat) -> 'c shape repr
end
\end{lstlisting}
\end{adjustbox}
\myhack{\vspace{-20pt}}
\caption{Tagless Final Representation of Core \polyjoin{}.}\label{polyjoins:fig:patternsymantics}
\end{figure}

%%% Local Variables:
%%% mode: latex
%%% TeX-master: "report"
%%% End:

\begin{figure}
  \centering
\begin{SubFloat}{\label{polyjoins:fig:hlistfunctor:signature}}%
\begin{minipage}[t][120pt]{.5\linewidth}
\begin{adjustwidth}{1ex}{}%
\begin{adjustbox}{scale=0.9,valign=T,left}%
\begin{lstlisting}
module type ?Hl? = sig
  type 'a el
  type _ hlist =
    | Z : unit hlist
    | S : 'a el * 'b hlist
             -> ('a * 'b) hlist
end
\end{lstlisting}%
\end{adjustbox}%
\end{adjustwidth}%
\end{minipage}%
\end{SubFloat}%
\begin{SubFloat}{\label{polyjoins:fig:hlistfunctor:functor}}%
\begin{minipage}[t][120pt]{.5\linewidth}
\begin{adjustwidth}{1ex}{}
\begin{adjustbox}{scale=0.9,valign=T,left}%
\begin{lstlisting}
module ?HList?(?E?: sig type 'a t end) =
struct
  type 'a el = 'a ?E?.t
  type _ hlist =
    | Z : unit hlist
    | S : 'a el * 'b hlist (*@\label{polyjoins:lst:hlistfunctor:cons-type}@*)
             -> ('a * 'b) hlist
  let nil = Z
  let cons h t = S (h,t)
end
\end{lstlisting}%
\end{adjustbox}%
\end{adjustwidth}%
\end{minipage}
\end{SubFloat}%
\myhack{\vspace{-10pt}}\caption{Heterogeneous Lists Definition.}
\end{figure}

%%% Local Variables:
%%% mode: latex
%%% TeX-master: "report"
%%% End:

\begin{figure}
  \centering
\begin{adjustbox}{valign=T,scale=.9,left,margin=7pt 0pt 0pt 0pt}
\begin{lstlisting}[aboveskip=0pt,belowskip=0pt]
module ?StdContext?(?T?: sig type 'a repr type 'a shape end) = struct
  open ?T?
  type _ var = Bind: 'a shape repr -> 'a var
  module ?Ctx? = ?HList?(struct type 'a t = 'a var end)
  type (_,_) shape = (* Shape translation judgment *) (*@\label{polyjoins:lst:concrete-context:shape:start}@*)
    | Base: (unit, unit) shape
    | Step: ('s, 'a) jsig  -> ('t * 's, 't repr * 'a) shape (*@\label{polyjoins:lst:concrete-context:shape:end}@*)
  (* Implementation of Context Formation (Figures (*@\ref{polyjoins:fig:polyjoin-core}@*) and (*@\ref{polyjoins:fig:patternsymantics}@*)) *)
  type ('a,'b) ctx = ('a,'b) shape * 'a ?Ctx?.hlist (*@\label{polyjoins:lst:concrete-context:ctx}@*)
  let from = fun s -> Bind s
  let cnil = (Base, ?Ctx?.nil)
  let (@.): type a c d. a var -> (c, d) ctx -> (a * c, a repr * d) ctx
      = fun v (shape, ctx) -> (Step shape, ?Ctx?.cons v ctx)
end
\end{lstlisting}
\end{adjustbox}\myhack{\vspace{-18pt}}
\caption{Default Variable Context Representation with Heterogeneous Lists.}\label{polyjoins:fig:concrete-context}
\end{figure}
%%% Local Variables:
%%% mode: latex
%%% TeX-master: "report"
%%% End:

\begin{figure}
  \centering
\begin{adjustbox}{left,valign=T,scale=.9}
\begin{lstlisting}[aboveskip=0pt,belowskip=0pt]
module ?MonadL? = struct
  type 'a repr  = 'a
  type 'a shape = 'a list
  type 'a pat   = 'a list
  (* Contexts: cf. Figure (*@\ref{polyjoins:fig:concrete-context}@*): *)
  include ?StdContext?(struct type 'a repr = 'a type 'a shape = 'a list)
  let where b p = if b then p else []
  let yield v = [v]
  let pair: 'a repr -> 'b repr -> ('a * 'b) repr = fun a b -> (a,b) (*@\label{polyjoins:lst:monadic-cartesian:pair}@*)
  (* Nested monadic bind by induction over the context derivation: *)
  let rec cart : type a b c. (a,b) ctx -> (a -> c list) -> c list = fun ctx k -> (*@\label{polyjoins:lst:monadic-cartesian:cart:start}@*)
    match ctx with
    | (Base, ?Ctx?.Z) -> k ()
    | (Step n, ?Ctx?.S (Bind ls, hs)) ->
       bind (fun x -> cart (n,hs) (fun xs -> k (x, xs))) (*@\label{polyjoins:lst:monadic-cartesian:cart:end}@*)
  let join: ('a, 'b) ctx -> ('b -> 'c pat) -> 'c shape repr = (*@\label{polyjoins:lst:monadic-cartesian:join}@*)
    fun ctx body -> cart ctx body
end
\end{lstlisting}
\end{adjustbox}
\myhack{\vspace{-20pt}}
\caption{Sequential Cartesian Product in \polyjoin{}.\todo{generalize to any monad}}\label{polyjoins:fig:monadic-cartesian}
\end{figure}

%%% Local Variables:
%%% mode: latex
%%% TeX-master: "report"
%%% End:

Here, we develop \polyjoin{} as a tagless final DSL. We make a simplification to
focus on the core ideas: event patterns do not expose timing (\eg, \lstinline|within| in
\Cref{polyjoins:fig:fireexampleocaml}) or other implicit metadata on events. We address these
features in \Cref{polyjoins:sec:case-study:-cart}.

\Cref{polyjoins:fig:polyjoin-core} defines the core syntax and typing rules of \polyjoin{} in natural
deduction style and \Cref{polyjoins:fig:patternsymantics} the corresponding tagless encoding as a
OCaml module signature.

\subsubsection{Expressions and Patterns} As before, we define syntax/typing rules $\hastype{M}{A}$
for the syntactic sort of expressions (\eg, \Cref{polyjoins:fig:tagless-final-primer:rules}) and
introduce a new sort for \emph{patterns}, $\formspat{P}{A}$,
meaning that pattern $P$
yields events of type $A$.
In this language, patterns consist only of \lstinline|where| (constraints/filter) and
\lstinline|yield| (lift expression of type $A$
to an event of type $A$).
However, we can always extend the language with more pattern forms, as needed.

\subsubsection{The essence of polyvariadic joins}
Rule \inflabel{join} formalizes the essence of
polyvariadic $n$-join
expressions, which we exemplified in \Cref{polyjoins:sec:fire-alarm-revisited}. A join expression requires a
valid pattern context $\Pi$
consisting of \lstinline|from| bindings. In the premise, context formation $\formsctx{\Pi}{\vec{A}}$
certifies that $\Pi$
is well-formed, exposing at the type-level (of the host language OCaml!) that $\Pi$
has the \emph{shape} $\vec{A}$,
which is an ordered, \emph{heterogeneous} sequence of types describing the number and element types
of the bindings in $\Pi$.
Keeping statically track of this information is crucial for programming polyvariadic definitions and
performing type-level computations.

As we motivated in \Cref{polyjoins:sec:joins-type-theor}, there
is a type-level functional dependency between the context of \lstinline|from| bindings and the
number and type of pattern variables in joins. Formally, context shape $\vec{A}$
translates to a context shape $\vec{B}$,
written $\pvars{\vec{A}}{\vec{B}}$.
The latter shape describes the number and types of available pattern variables, which are used in
the rightmost premise of rule \inflabel{join}. This premise defines the body of the join pattern.
Its variables are assumptions of a derivation ending in a pattern form $P$. The derivation generalizes
the usual implication introduction rule of natural deduction to $n$
assumptions and is represented in OCaml as an uncurried, $n$-ary
function value/HOAS binding.  Intuitively, the body $P$
of the join pattern defines how to construct a single output element of type $C$,
from elements $\vec{x}$
extracted from the join's input sources.  As a result, the type of the whole join expression lifts
this specification for single $C$
elements into a whole collection $\Type{Shape}[C]$.
Here, the abstract type constructor $\Type{Shape}[\cdot]$
corresponds to the type constructor $S[\cdot]$ in \Cref{polyjoins:def:joins}. Overall, the \lstinline|join|
form defines an $n$-ary join followed by a map
$$\overrightarrow{\Type{Shape}[A]}\Rightarrow\Type{Shape}[\vec{A}]\Rightarrow\Type{Shape}[C]$$ in the sense
of our earlier definition.

\subsubsection{Context Formation} In contrast to standard treatments of variable bindings and
context, our pattern contexts $\Pi$
are nameless, \ie, they are \emph{not} sequences of variable/type pairs $\overrightarrow{x: A}$,
because we cannot directly model variable names in OCaml's type language. Instead, pattern contexts
$\Pi$
just witness the shape $\vec{A}$, which is enough to compute the appropriate signature of a join pattern.

Variable introduction $\isvar{V}{A}$
witnesses that a binding term $V$
introduces an anonymous variable of type $A$.
In core \polyjoin{}, only \lstinline|from|-bindings can introduce a variable via rule
\inflabel{from}.  The latter establishes that an anonymous variable of type $A$
must come from an input source term $M$ of $\Type{Shape}[A]$. The rules for context formation $\formsctx{\Pi}{\vec{A}}$
specify that contexts are inductively formed by the terms \lstinline|cnil| (empty context) and
\lstinline|@.| (prepending of variable to context).

\subsubsection{Shape Translation} The purpose of this judgment is controlling how the types of pattern variables
are computed from the shape of the pattern context $\Pi$. The core version of \polyjoin{}
trivially establishes $\vec{B} = \vec{A}$, \ie, the $i$th
pattern variable binds a DSL term representing an element of the type $A_i$. In
\Cref{polyjoins:sec:case-study:-cart}, we will change the rules for this judgment, attaching timing
information to pattern variables.

\subsubsection{OCaml Representation of PolyJoin}\label{polyjoins:sec:ocaml-repr-polyj}
Following the principles of \Cref{polyjoins:sec:primer-tagless-final}, we define the
intrinsically-typed syntax of \polyjoin{} in an OCaml module signature
(\Cref{polyjoins:fig:patternsymantics}).  Each of the judgment forms presented here corresponds to
an abstract type constructor (which we marked in the comments) and each rule to a function.  Note
that the OCaml version bundles context formation and shape translation into a single type
constructor/judgment. This way, we avoid requiring the user to manually supply the derivation of the
shape translation judgment, since it can be inductively defined from the structure of the context
formation. Accordingly, the context formation rules \lstinline|cnil| and \lstinline|@.| also compute
the shape translation in the second type parameter of \lstinline|ctx|.

Indeed, given this module signature, the unification-based Hindley-Milner (HM) type system of OCaml
is sufficient for computing the correct type of the polyvariadic \lstinline|join| form, for any
expressible pattern context.  For example, a partial application with three bindings
\begin{lstlisting}[numbers=none]
join ((from source1) @. (from source2) @. (from source3) @. cnil)
\end{lstlisting}
yields a three-ary pattern abstraction
\begin{lstlisting}[numbers=none]
((a1 repr * (a2 repr * (a3 repr * unit)) -> '_b pat) -> '_b shape repr
\end{lstlisting}
as intended, given that \lstinline|source$i$|, has type \lstinline|a$i$ shape repr|, for
$i\in\Set{1,2,3}$. We model heterogeneous sequences via nested binary products at
the type-level.

\subsection{Polyvariadic Programming}\label{polyjoins:sec:progr-with-polyv}

Our tagless final embedding of \lstinline|join|
(\Cref{polyjoins:fig:patternsymantics}) makes polyvariadicity explicit in its signature, by quantifying over all
possible variable contexts \lstinline|('a,'b)$\ $ctx|.
Next, we address how to program against such \emph{context-polymorphic} interfaces,
in order to implement concrete interpreters of \lstinline|?Symantics?| in \Cref{polyjoins:fig:patternsymantics}.

\subsubsection{Heterogeneous Sequences}\label{polyjoins:sec:abstr-over-heter}

We program against context-polymorphic interfaces with generalized algebraic data types
(GADTs)~\cite{Johann:2008aa,Cheney:2003aa}, using heterogeneous lists~\cite{Kiselyov:2004aa}
\begin{lstlisting}[numbers=none,aboveskip=1pt,belowskip=1pt]
type _ hlist = Z : unit hlist
             | S : 'a * 'b hlist -> ('a * 'b) hlist
\end{lstlisting}
One can tell that \lstinline|hlist| is a GADT from
the underscore in the type parameter position.  The constructors \lstinline|Z| (empty list) and
\lstinline|S| (list cons) constrain the shape of the possible types that can be filled into the type
parameter, as a form of bounded polymorphism.  In the case of \lstinline|hlist|, the type parameter
describes the exact shape of a heterogeneous list value. For example, the value
\begin{lstlisting}[numbers=none]
S (1, S ("two", S (3.0, Z)))
\end{lstlisting}
has type \lstinline|(int * (string * (float * unit)))$\ $hlist|. GADTs enable the compiler to prove more precise properties of programs.
A classic example is the safe head function on \lstinline|hlist|
\begin{lstlisting}[numbers=none,aboveskip=0pt,belowskip=0pt]
let safe_head: type a b. (a * b) hlist -> a =
  function S (x,_) -> x
\end{lstlisting}
which statically guarantees that it can be only invoked with non-empty list arguments.
This works because we constrain the shape of the argument's type parameter to
\lstinline|(a * b)$\ $hlist|.
Hence, by the definition of \lstinline|hlist| above, the argument can only be of the form
\lstinline|S (x,_)|.

\subsubsection{Uniform Heterogeneity}\label{polyjoins:sec:constr-heter} We sometimes require stronger
invariants on the heterogeneous context/list shape certifying that elements are uniformly enclosed
in a given type constructor.  For example, having a heterogeneous list of homogeneous lists:
\lstinline|int list * (string list * (float list * unit))|.  It is hard to enforce such invariants
using only the bare \lstinline|hlist| type above, because the type parameters of elements
\lstinline|'a| in the \lstinline|S| constructor quantifies over any type.
We overcome this issue by defining a \emph{family} of constrained \lstinline|hlist| types,
which is parameteric over a type constructor \lstinline|'a el|, applied element-wise (\Cref{polyjoins:fig:hlistfunctor:signature}).
We may create concrete instances of constrained \lstinline|hlist|s
in form of OCaml modules, with the functor \lstinline|?HList?| (\Cref{polyjoins:fig:hlistfunctor:functor}).
For example, the modules
\begin{lstlisting}[numbers=none,aboveskip=1pt,belowskip=1pt]
module ?HL? = ?HList?(struct type 'a t = 'a end)         (* Standard hlist *)
module ?Lists? = ?HList?(struct type 'a t = 'a list end) (* hlist of lists *)
\end{lstlisting}
define unconstrained \lstinline|hlist|s and \lstinline|hlist|s of homogeneous lists, respectively.
To distinguish between concrete \lstinline|hlist| variations in programs, we
qualify the types and constructors by the defining module's name, \eg, the function
\begin{lstlisting}[aboveskip=1pt,belowskip=1pt]
let rec enclose: type a. a ?HL?.hlist -> a ?Lists?.hlist =
  function
  | HL.Z -> ?Lists?.nil
  | HL.S (hd,tl) -> ?Lists?.(cons [hd] (enclose tl))
\end{lstlisting}
is a polyvariadic function that element-wise converts unconstrained \lstinline|hlist|s into
\lstinline|hlist|s of homogeneous lists.

\subsubsection{Shape Preservation}\label{polyjoins:sec:shape-preservation}
The definition of \lstinline|enclose| above features a subtle, but important design principle.  Its
signature \lstinline|type a. a ?HL?.hlist -> a ?Lists?.hlist| is polyvariadic, because it
quantifies\emph{over all possible shapes} \lstinline|a| of \lstinline|hlist|s. Notice that the same
shape \lstinline|a| also appears in the codomain of the function type. That means,
\lstinline|enclose| is \emph{shape preserving}, \ie, it does not change the number of elements or
their basic types. But it changes the enclosing type constructor.  The invariance of the type
parameter \lstinline|a| is a way of encoding the relation between heterogeneous input and output in
OCaml's type system, without the need of advanced type-level computation. From the definition of the
\lstinline|HL| and \lstinline|Lists| modules and the invariance in shape \lstinline|a|, the above
signature certifies that for all $n\geq 0$,
\lstinline|enclose| takes $n$-tuples
\lstinline|(a$_1$,...,a$_n$)| to $n$-tuples \lstinline|(a$_1$ list,...,a$_n$ list)|.
Shape preservation is crucial to compose independently developed polyvariadic components in
a type-safe manner.

\subsubsection{Contexts as Heterogeneous Lists}
Interpreters of the join form
\begin{lstlisting}[numbers=none,aboveskip=1pt,belowskip=1pt,xleftmargin=70pt]
join: ('a, 'b) ctx -> ('b -> 'c pat) -> 'c shape repr
\end{lstlisting}
must program against precisely this type signature. Somehow, the tagless interpreter should be able
to \emph{extract} individual element values \lstinline|a1 repr|\ldots\lstinline|an repr| (type
parameter \lstinline|'b|) from $n$
input shapes having type \lstinline|a1 shape repr| \ldots \lstinline|'an shape repr| and bind them
to the pattern body \lstinline|('b -> 'c pat)|. Where do they come from? The formal system in
\Cref{polyjoins:fig:polyjoin-core} specifies that the derivation of the context parameter
\lstinline|('a, 'b)$\ $ctx|
includes the bound input shapes, by rule \inflabel{from}. An interpreter should be able to
analyze how the judgment \lstinline|('a, 'b)$\ $ctx|
was derived, to get ahold of the input shapes.  Hence, interpreters of \lstinline|join| really
are proofs by induction over the context derivation, which we can express in OCaml using
a recursive function over a GADT representation of the derivation. GADTs enable case analyses
over derivations.

Therefore, we instantiate the abstract context signature (\Cref{polyjoins:fig:patternsymantics}) in
terms of constrained heterogeneous lists and other GADTs, as shown in
\Cref{polyjoins:fig:concrete-context}.  The variable representation \lstinline|'a var| becomes a
GADT, whose constructor \lstinline|Bind| encloses an input source representation and links it to an
element variable. The context representation \lstinline|'a ctx| is instantiated to the constrained
heterogeneous list type \lstinline|?Ctx?.hlist|, which stores \lstinline|Bind| values. Thus, contexts
are concrete data values that the implementation of the interpreter is free to inspect and
manipulate.  We also need to model the shape translation judgment from
\Cref{polyjoins:fig:polyjoin-core}, which is codified by the \lstinline|('a, 'b)$\ $shape|
GADT (\Cref{polyjoins:fig:concrete-context},
Lines~\ref{polyjoins:lst:concrete-context:shape:start}-\ref{polyjoins:lst:concrete-context:shape:end}).
As explained in \Cref{polyjoins:sec:ocaml-repr-polyj}, context formation and shape translation are
combined into a single judgment, which translates to their product in
\Cref{polyjoins:fig:concrete-context}, Line~\ref{polyjoins:lst:concrete-context:ctx}.

\subsubsection{Example: Sequential Cartesian Product}\label{polyjoins:sec:impl-cust-bind}
We conclude this section with an example \polyjoin{} interpreter, implementing an old friend: the
nested monadic translation from LINQ/comprehensions (\Cref{polyjoins:sec:motivation}) over lists, yielding a cartesian
product. \Cref{polyjoins:fig:monadic-cartesian} shows the full implementation.\footnote{For the
  example, we extended the DSL with support for constructing binary pairs
  (\Cref{polyjoins:fig:monadic-cartesian}, Line~\ref{polyjoins:lst:monadic-cartesian:pair}).} The
interpreter is meta-circular, interpreting expressions as OCaml values and works with lists as input
shape representation. In this case, we also set the pattern type to lists.
\begin{lstlisting}[aboveskip=1pt,belowskip=1pt]
open ?MonadL?
join ((from [1]) @. (from ["2";"3"]) @. (from [4.0;5.0]) @. cnil)
     (fun (x, (y, (z,()))) -> (yield (pair x (pair y z))))
(* result: (int * (string * float)) list =
   [(1, ("2", 4.)); (1, ("2", 5.)); (1, ("3", 4.)); (1, ("3", 5.))] *)
\end{lstlisting}
The tagless interpreter \lstinline|MonadL| implements the $n$-nary
nested monadic bind, by a straightforward structural induction (see above) over the standard context
representation (\Cref{polyjoins:fig:concrete-context}) in the function \lstinline|cart|
(\Cref{polyjoins:fig:monadic-cartesian},
Line~\ref{polyjoins:lst:monadic-cartesian:cart:start}-\ref{polyjoins:lst:monadic-cartesian:cart:end}).
This function encloses a given continuation function \lstinline|k| of arity $n$ with $n$ layers of monadic list bindings.
The $n$-ary body of the join pattern is the innermost layer of this nesting (Line~\ref{polyjoins:lst:monadic-cartesian:join}).

In the next section, we show that \polyjoin{} supports more than sequential variable bindings, which
go beyond LINQ and standard comprehension implementations.

%%% Local Variables:
%%% mode: latex
%%% TeX-master: "report"
%%% End:

% !TEX root = ./report.tex
\section{Case Study: Event Correlation with Algebraic Effects}\label{polyjoins:sec:case-study:-cart}

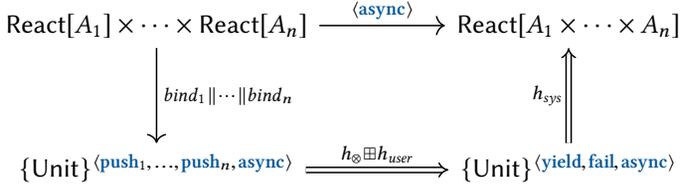
\begin{figure}
  \centering
  $$\begin{tikzcd}[column sep=huge,row sep=huge]
    \Type{React}[A_1]\times \cdots \times \Type{React}[A_n] \arrow{r}{\anglep{\eff{async}}}  \arrow[rightarrow]{d}{{\mathit{bind}_1\parallel\cdots\parallel\mathit{bind}_n}}  & \Type{React}[A_1\times\cdots\times A_n] \\
          \thunk{\Type{Unit}}^{\anglep{\eff{push}_1,\ldots, \eff{push}_n,\eff{async}}} \arrow[Rightarrow]{r}{h_{\otimes}\hcomp h_{\mathit{user}}} & \thunk{\Type{Unit}}^{\anglep{\eff{yield},\eff{fail},\eff{async}}}  \arrow[Rightarrow]{u}{h_{\mathit{sys}}}
  %\Type{React}[A_1]\times \cdots \times \Type{React}[A_n] \arrow[r, "\anglep{\eff{async}}"] \arrow[d, dashed, rightarrow, "\text{bind}"]
  %& \Type{React}[A_1\times\cdots\times A_n]  \arrow[d,  Leftarrow, "h_{\mathit{sys}}"] \\
  %\thunk{\Type{Unit}}^{\anglep{\eff{push}_1,\ldots, \eff{push}_n,\eff{async}}} \arrow[r, Rightarrow, "h_{\otimes}\hcomp h_{\mathit{user}}"] & \thunk{\Type{Unit}}^{\anglep{\eff{yield},\eff{fail},\eff{async}}}
\end{tikzcd}$$
\myhack{\vspace{-4ex}}
\caption{Join Computations in \corrl{}.}\label{polyjoins:fig:cartesius-overview}
\end{figure}

%%% Local Variables:
%%% mode: latex
%%% TeX-master: "report"
%%% End:

In this section, we embed
the \corrl{} language by \citet{Bracevac:2018aa} with \polyjoin{}. \corrl{} is a general
model for event correlation over asynchronous streams that is based upon algebraic effect
handlers~\cite{Plotkin:2003aa,Plotkin:2009aa}.
Bra\v{c}evac et al.\ outline a macro translation of their event pattern syntax into effects and
handlers. However, their prototype implementation lacks this syntax and is not polyvariadic,
supporting only a handful of hard-coded arities.  \polyjoin{} enables the first fully polyvariadic
implementation with proper pattern syntax embedding, in the multicore OCaml
dialect~\cite{dolan2017concurrent}.\footnote{The full implementation is available at
  \implementationurl{}.} \corrl{} is an
interesting larger example, because it has event patterns that expose event timing and concurrent
binding semantics.  Another important point of the example is to showcase how to
program a polyvariadic backend against \polyjoin{}'s polyvariadic interfaces.

\subsection{Overview of \corrl{}}\label{polyjoins:sec:cartesius-language}
The diagram in
\Cref{polyjoins:fig:cartesius-overview} illustrates how polyvariadic join computations in the sense
of \Cref{polyjoins:def:joins} (top row) relate to \corrl{}' effect-based representation (bottom
row).  In this setting, function types ($\to$)
are annotated with Koka-style effects rows~\cite{Leijen:2017aa}, indicating the side effect a
function call may induce. We write concrete effect types in blue font. \corrl{} joins perform
global asynchrony effects $\anglep{\eff{async}}$
(cf.~\cite{dolan2017concurrent,Leijen:2017ab}), accounting for the inversion of control present in
event correlation (\Cref{polyjoins:sec:join-example}). The other kind of arrow
($\Rightarrow$)
indicates \emph{effect handlers}, which transform effectful computations into other effectful
computations.  The overall idea is that \emph{event notifications} are effects, \emph{event sources}
are computations inducing the effects and effect handlers are \emph{event observers}.

\corrl{} combines $n$ input sources into a unit-valued effectful computation of type
$$\thunk{\Type{Unit}}^{\anglep{\eff{push}_1,\ldots, \eff{push}_n,\eff{async}}}$$
by a \emph{parallel composition of bindings} (left column). We write computation types in curly
braces (\ie, they designate thunks) annotated with a row specifying the possible side effects.
Intuitively, this computation is an asynchronous process that in parallel subscribes to $n$
input event sources, interleaving and forwarding their event notifications.  We model these
notifications in terms of a family of $n$
heterogeneous effect operations $(\eff{push}_i)_{1\leq i \leq n}$,
which correspond to a sequence of effect declarations in multicore OCaml, informally written:
\begin{lstlisting}[numbers=none,mathescape=true,xleftmargin=120pt,aboveskip=1pt,belowskip=1pt]
(effect &@Push&@$_i$: 'a$_i$ -> unit)$_{1\leq i \leq n}$
\end{lstlisting}
That is, effects are named operations in the algebraic effect setting, having the signature of
functions.  Computations invoke them to induce the effect.  Here, \lstinline|&@Push&@$_i$|
carries in its parameter a typed event value from the $i$-th
event source.

The effect row statically certifies that the binding is indeed parallel: in contrast to sequential
monadic binding (cf.~\Cref{polyjoins:sec:event-patterns-as}) there is no control dependency between
the event notification effects, \ie,
$${\anglep{\eff{push}_1,\ldots, \eff{push}_n,\eff{async}}}\equiv {\anglep{\eff{push}_{\pi(1)},\ldots, \eff{push}_{\pi(n)},\eff{async}}}$$
are equivalent in the effect type system, for any permutation $\pi: \Set{1\ldots n}\to\Set{1\ldots n}$,
certifying that the event notifications may occur in any order and arbitrarily often.

Invocations of effect operations are discharged by effect handlers, intuitively generalizations of
exception handlers, which can resume back (similarly to coroutines~\cite{Moura:2009aa}).
Handlers may induce additional effects while discharging the observed effects of the underyling
computation. Hence, effect handlers define a custom semantics/implementation of effect operations and
are transformations ($\Rightarrow$) between effectful computations.

The bottom row of \Cref{polyjoins:fig:cartesius-overview} indicates that an effect handler
$h_{\otimes}\hcomp h_{\mathit{user}}$
transforms the above event-observing process into a process that generates and tests correlated
$n$-tuples, either yielding or discarding them through effects:
\begin{lstlisting}[numbers=none,xleftmargin=70pt,aboveskip=1pt,belowskip=1pt]
effect &@Yield&@: 'a$_{1}\times\cdots\times\,$'a$_n$ -> unit (* Output tuple *)
effect &@Fail&@:  unit -> 'a           $\;$(* Discard tuple *)
\end{lstlisting}
The handler $h_{\otimes}\hcomp h_{\mathit{user}}$
is a composition ($\hcomp$)
of a system default handler $h_{\otimes}$
and a user-defined handler $h_{\mathit{user}}$,
which implements the actual event correlation logic. Intuitively, the correlation logic is an
$n$-way
coroutine that coordinates the $n$
event subscriptions over the event sources.  Finally, the handler $h_{\mathit{sys}}$
is part of the system runtime and sends generated $n$-tuples
into the output event source by handling the $\eff{yield}$/$\eff{fail}$
effects.

\subsubsection{Callback Logic}\label{polyjoins:sec:n-entangl-cont}
As part of the join computation, the system handler $h_\otimes$
in \Cref{polyjoins:fig:cartesius-overview} stores observed event notifications locally
in $n$ heterogeneous mailboxes. Each new event notification triggers a generic cartesian
product computation over these mailboxes, for generating and testing
$n$-tuples that satisfy the join's constraints.

Due to the asynchrony of the $n$
event sources, the above strategy requires registering $n$
heterogeneous callbacks/continuation functions $(\kappa_i)_{1\leq i \leq n}$,
which together implement the cartesian product. In more formal terms, the callbacks have the
shape depicted in \Cref{polyjoins:fig:focusing},
where $M[\cdot]$ is a collection type for mailboxes and $\otimes$ a binary cartesian product
operator on mailboxes.
Each callback $\kappa_i$ represents the behavior when the $i$-th event source fires its next event,
which is bound to the $\textcolor{red}{x}$ variable.
\ie, it replaces the $i$-th mailbox with the singleton mailbox $\textcolor{red}{\anglep{x}}$
in the cartesian product over all mailboxes $m_i:M[A_i]$ and computes it.

Writing callbacks for joining asynchronous computations leads to another manifestation of the
binding problem for event correlation (cf.~\Cref{polyjoins:sec:event-patterns-as}), this time in
terms of the well-known continuation monad.\footnote{It is common practice in mainstream asynchronous
programming libraries (\eg, \csharp{}~\cite{Bierman:2012aa} or Scala~\cite{EPFL:2013aa}) to join
multiple computations by nesting callback subscriptions, which introduces unnecessary sequential
dependencies.}  To avoid these problems, the $n$
callbacks perform a ``focusing in the middle'' of the expression $m_1\otimes\cdots \otimes m_{n}$
for each possible position. The possible focus positions reflect the choices that the external
environment can make to supply events (inversion of control).

\subsubsection{Restriction Handlers}\label{polyjoins:sec:restriction-handlers}

\citet{Bracevac:2018aa} use pattern syntax similar to \polyjoin{}. In addition, their syntax
supports a way to compose and inject user-defined effect handlers (the $h_{\mathit{user}}$
component in \Cref{polyjoins:fig:cartesius-overview}), called \emph{restriction handlers}. They
model composable sub-computations that change the behavior of the whole computation.
Without custom restrictions, the variable bindings in \corrl{} join patterns have
a universal quantification semantics, executing the cartesian product above.

\Cref{polyjoins:fig:cartesius-combine-latest} shows a preview of our \polyjoin{} embedding for
\corrl{}.  The pattern specifies that the \lstinline|most_recently| restriction handler shall apply
to the first (\lstinline|p0|) and second (\lstinline|p1|) event source of the join
(\Cref{polyjoins:lst:cartesius-combine-latest:mostrecently}). Or equivalently: the first and second
pattern variable should have the most recently binding semantics.  For now, we leave \lstinline|p0|,
\lstinline|p1| underspecified, they are essentially nameless
indices~\cite{De-Bruijn:1972aa} into the context of the join.\footnote{\citet{Bracevac:2018aa} apply
  the restrictions to the event source \emph{values} in their pen and paper examples. However,
  this is technically inaccurate, because their formal semantics defines the restriction handlers
  in the nameless form. The \polyjoin{} version of the syntax accurately reflects that restrictions
  refer to input positions in the context of the join.}
Intuitively, the operator \lstinline@|++|@ represents effect handler composition $\hcomp$.

These restriction handlers change the behavior of the default cartesian product to achieve the
\emph{combine latest} correlation semantics, which is not expressible with nested joins
(cf.~\Cref{polyjoins:sec:limit-linqs-encod}). For example, from the input
\begin{lstlisting}[numbers=none,language=poorman,columns=fixed,aboveskip=1pt,belowskip=1pt]
temp_sensor  = < (120,  1), (50,    3), (20,   5) >
smoke_sensor = < (true, 1), (false, 2), (true, 4) >
\end{lstlisting}
we get
\begin{lstlisting}[numbers=none,language=poorman,columns=fixed,aboveskip=1pt,belowskip=1pt]
< ((120, true), [1,1]), ((120, false), [1,2]), ((50, false), [2,3]),
  ((50,  true), [3,4]), ((20,  true),  [4,5]) >$.$
\end{lstlisting}
That is, the correlation always reflects the most up to date input events.

The restriction handler mechanism enables changing the variable binding semantics of the correlation
at the level of individual variables, in contrast to \Cref{polyjoins:sec:solution}, where we
had a polymorphic, but \emph{uniform} variable binding semantics for all $n$ pattern variables.
For example, if we leave out the \lstinline|most_recently p1| restriction
in \Cref{polyjoins:lst:cartesius-combine-latest:mostrecently}, then the second variable binding
would retain the default cartesian product semantics: ``join the most up to date value of
the first event source with \emph{all} events from the second''.

Moreover, \corrl{} features restriction handlers constraining more than one variable binding
simultaneously. For example, if we replaced
\Cref{polyjoins:lst:cartesius-combine-latest:mostrecently} with the restriction handler
\lstinline|aligning p0 p1|, then we would obtain a correlation that \emph{zips} its inputs
(cf.~\cite{Bracevac:2018aa}):
\begin{lstlisting}[numbers=none,language=poorman,columns=fixed,aboveskip=1pt,belowskip=1pt]
< ((120,true),  [1,1]), ((50,false), [2,3]), ((20,true),  [4,5]) >$.$
\end{lstlisting}
The general form of this restriction is \lstinline|aligning $K$|, \ie,
a given \emph{subset} $K \subseteq \Set{1\ldots n}$ of inputs should be
aligned.

\subsubsection{Conclusion and Challenges}\label{polyjoins:sec:cartesius-overview:summary}
The \corrl{} design introduces additional (polyvariadic) syntax elements and sub-components, which
we need to address for a statically type-safe embedding and implementation of the language.
We will show that these are well within reach of \polyjoin{}'s conceptual tools.

\myparagraph{Polyvariadic Effect Declarations and Handlers}
The effect declarations of \corrl{}  have
type signatures which are functionally dependent on the types of event sources to be joined.
Yet, no linguistic concept in the language of \emph{declarations}
allows calculating a sequence of heterogeneous effect declarations from the types of inputs.
This is a new instance of polyvariadicity, because so far, we addressed it in the
type and expression languages only. Another issue is how to idiomatically write polyvariadic
handlers for these effects.

\myparagraph{Event Meta Data} \corrl{} patterns expose time data of events and accordingly
constraints on time data, \eg, \lstinline|within| in \Cref{polyjoins:fig:fireexampleocaml}.

\myparagraph{The Focusing Continuations Problem} The family $(\kappa_i)_{1 \leq i \leq n }$
of callback functions is a non-trivial polyvariadic definition.  The issue is calculating a
polyvariadic representation of the different focusing positions and replacements in the
heterogeneously typed sequence $$(m_i)_{1\leq i \leq n}: M[A_1]\times\cdots\times M[A_n]$$
of mailboxes.  Seemingly, we require a non-trivial type-level computation to represent the focusing.
\Ie, given the mailbox sequence type, calculate ($\leadsto$) the type of all possible ways to ``punch holes''
into the mailbox sequence:
$$ M[A_1]\times\cdots\times M[A_n] \leadsto (M[A_1]\times\cdots\times M[A_{i-1}]\times\textcolor{ACMRed}{[\;\cdot\; ]}\times M[A_{i+1}]\times\cdots\times M[A_n] )_{1\leq i\leq n}.  $$
This type is effectively the zipper~\cite{Huet:1997aa} over the mailbox sequence type.  Nevertheless, this
problem is worthwhile solving, because it is of general interest for asynchrony and concurrency
implementations in statically-typed sequential languages.

\myparagraph{Safe, Reusable, Modular Restriction Handlers} Is important that users should not be able to
specify ill-defined restrictions in a join pattern. \Eg, providing a De Bruijn index that refers to
a non-existent position in the context or supplying a $K$
which is not a subset of $\Set{1\ldots n}$
in the case of \lstinline|aligning $K$|.
It is also important that the implementer of restrictions can avoid repetition and provide
statically safe, yet maximally reusable definitions. \Eg, if the \lstinline|aligning $K$|
value is compatible in the context of an $n$-way
join, then it should be compatible for an ($n+1$)-way
join, too.  Finally, to foster extensibility and modularity, new kinds of restriction handlers
should be programmable separately and independently from concrete join computations.

\begin{figure}
\syntaxcategorynf{Expressions and Patterns}{\fbox{$\hastype{M}{A}$}  \fbox{$\formspatm{\vec{C}}{P}{A}$}}
\begin{mathpar}
\inferrule[where]{\hastype{M}{\Type{Bool}}\\ \formspatm{\vec{C}}{P}{A}}{\formspatm{\vec{C}}{\mathtt{where}\;M\;P}{A}}
\and
\inferrule[yield]{\hastype{M}{A}}{\formspatm{\vec{C}}{\mathtt{yield}\;M}{A}}
\and
\inferrule[mmerge]{\hastype{M}{\Type{Meta}}\\\hastype{N}{\Type{Meta}}}{\hastype{M\sqcup N}{\Type{Meta}}}
\and
\inferrule*[Right=join]{\formsctx{\Pi}{\vec{A}}\\ \pvars{\vec{A}}{\vec{B}}\\\formsext{\vec{A}}{X}\\\inferrule*{\inferrule*[vdots=1em]{}{[\overrightarrow{\hastype{x}{B}}]}}{\formspatm{\vec{A}}{P}{C}}}{\hastype{\mathtt{join}\;\Pi\;X\;(\vec{x}.P)}{\Type{Shape}[C]}}
\end{mathpar}
%\syntaxcategorynf{Context Formation}{\fbox{$\isvar{V}{A}$} \fbox{$\formsctx{\Pi}{\vec{A}}$}}
% \begin{mathpar}
%   \inferrule*[Right=from,vcenter]{\hastype{M}{\Type{Shape}[A]}}{\isvar{\mathtt{from}\; M}{A}}
%   \and
%   \inferrule*[Right=cnil,vcenter]{}{\formsctx{\mathtt{cnil}}{\Type{\varnothing}}}
%   \and
%   \inferrule*[Right=cat,vcenter]{\isvar{V}{B}\\ \formsctx{\Pi}{\vec{A}}}{\formsctx{V \mathbin{\mathtt{@.}}\Pi}{B,\vec{A}}}
% \end{mathpar}
\syntaxcategory{Shape Translation}{$\pvars{\vec{A}}{\vec{B}}$}\myhack{\vspace{-5ex}}
\begin{mathpar}
  \inferrule*[Right=tz,vcenter]{}{\pvars{\varnothing}{\varnothing}}\and
  \inferrule*[Right=ts,vcenter]{\pvars{\vec{A}}{\vec{B}}}{\pvars{C,\vec{A}}{(C\times\Type{Meta}),\vec{B}}}\and
\end{mathpar}
\syntaxcategorynf{Shape (Multi)Projection}{\fbox{$\ctxproj{n}{A}{\vec{B}}$} \fbox{$\ctxmproj{\vec{n}}{\vec{A}}{\vec{B}}$}}
\begin{mathpar}
  \inferrule*[Right=pz,vcenter]{}{\ctxproj{\mathtt{pz}}{A}{A,\vec{B}}}\and
  \inferrule*[Right=ps,vcenter]{\ctxproj{n}{A}{\vec{B}}}{\ctxproj{\mathtt{ps}\;n}{A}{C,\vec{B}}}\and
  \inferrule*[Right=mz,vcenter]{}{\ctxmproj{\mathtt{\anglep{}}}{\varnothing}{\vec{A}}}\and
  \inferrule*[Right=ms,vcenter]{\ctxproj{n}{C}{\vec{B}}\\\ctxmproj{\vec{m}}{\vec{A}}{\vec{B}}}{\ctxmproj{n,\vec{m}}{C,\vec{A}}{\vec{B}}}
\end{mathpar}
\syntaxcategory{Contextual Extension Operations}{$\formsext{\vec{A}}{X}$}\myhack{\vspace{-4ex}}
\begin{mathpar}
\inferrule*[vcenter,Right=eempty]{}{\formsext{\vec{A}}{\mathtt{enil}}}\and
\inferrule*[vcenter,Right=emerge]{\formsext{\vec{A}}{X}\\\formsext{\vec{A}}{Y}}{\formsext{\vec{A}}{X\mathbin{\mathtt{++}} Y}}
\end{mathpar}
\myhack{\vspace{-10pt}}\caption{\polyjoin{} with Meta Data and Contextual Extensions.}\label{polyjoins:fig:polyjoin-cartesius}
\end{figure}
%%% Local Variables:
%%% mode: latex
%%% TeX-master: "report"
%%% End:

\begin{figure}
  \centering
\begin{adjustbox}{left,margin=0pt 0pt 25pt 0pt,valign=T,scale=.9}%
\begin{lstlisting}[belowskip=0pt,aboveskip=0pt]
join ((from temp_sensor) @. (from smoke_sensor) @. cnil)
      ((most_recently p0) |++| (most_recently p1)) (*@\label{polyjoins:lst:cartesius-combine-latest:mostrecently}@*)
      (fun ((temp,t1), ((smoke,t2), ())) -> (*@\label{polyjoins:lst:cartesius-combine-latest:deep}@*)
        yield (pair temp  smoke))
\end{lstlisting}%
\end{adjustbox}%
\myhack{\vspace{-10pt}}
\caption{Example: Join Pattern with \corrl{}-style Contextual Extension.}\label{polyjoins:fig:cartesius-combine-latest}
\end{figure}
%%% Local Variables:
%%% mode: latex
%%% TeX-master: "report"
%%% End:

\begin{figure}
  \centering
\begin{minipage}[t]{.5\linewidth}
\begin{adjustwidth}{1ex}{}%
\begin{adjustbox}{valign=T,scale=.9}
\begin{lstlisting}[belowskip=0pt,aboveskip=0pt]
module type ?SLOT? = sig
  type t
  effect &@Push&@: t -> unit
  effect &@Get&@:  unit -> t mailbox (*@\label{polyjoins:lst:slotsignature:getset0}@*)
  effect &@Set&@:  t mailbox -> unit (*@\label{polyjoins:lst:slotsignature:getset1}@*)
end
type 'a slot =                   (*@\label{polyjoins:lst:slotsignature:typealias0}@*)
  (module ?SLOT? with type t = 'a) (*@\label{polyjoins:lst:slotsignature:typealias1}@*)
\end{lstlisting}%
\end{adjustbox}%
\end{adjustwidth}%
\end{minipage}%
\begin{minipage}[t]{.5\linewidth}
\begin{adjustbox}{valign=T,scale=.9}
\begin{lstlisting}[belowskip=0pt,aboveskip=0pt]
module type ?YF? = sig
  type t
  effect &@Yield&@: t -> unit
  effect &@Fail&@:  unit -> 'a
end
type 'a yieldfail =
  (module ?YF? with type t = 'a) (*@\label{polyjoins:lst:slotsignature:typealiasyf}@*)
\end{lstlisting}%
\end{adjustbox}%
\end{minipage}%
  \caption{Generative Effects from Modules in Multicore OCaml.}\label{polyjoins:fig:slotsignature}
\end{figure}

%%% Local Variables:
%%% mode: latex
%%% TeX-master: "report"
%%% End:

\begin{figure}
  \centering
\begin{adjustbox}{left,valign=T,scale=.9}
\begin{lstlisting}[aboveskip=0pt,belowskip=-10pt]
(* type a b. a Slots.hlist -> b handler *)
let memory slots = poly_handler slots (fun (s: (module ?SLOT?)) ->
  let module ?S? = (val s) in
  let mbox: ?S?.t mailbox ref = ref (mailbox ()) in (*@\label{polyjoins:fig:polyhandler:cell}@*)
    fun action -> try action () with
      | effect (?S?.&@Get&@ ()) k -> continue k !mbox
      | effect (?S?.&@Set&@ m)  k -> mbox := m; continue k ())
\end{lstlisting}
\end{adjustbox}\myhack{\vspace{-10pt}}
\caption{Example Polyvariadic State Handler.}\label{polyjoins:fig:polyhandler}
\end{figure}
%%% Local Variables:
%%% mode: latex
%%% TeX-master: "report"
%%% End:

\begin{figure}
  \begin{adjustbox}{valign=T,left,scale=.9}
\begin{lstlisting}[aboveskip=0pt,belowskip=5pt]
let where: bool repr -> ('c,'a) pat -> ('c,'a) pat =
  fun cond body meta yf ->
    if cond then (body meta yf) else fail_with yf
\end{lstlisting}
\end{adjustbox}
\begin{adjustbox}{valign=T,left,scale=.9}
\begin{lstlisting}[aboveskip=0pt,belowskip=-10pt]
let yield: 'a repr -> ('c,'a) pat =
  fun result meta yf -> (result, (merge_all meta))
\end{lstlisting}
  \end{adjustbox}\myhack{\vspace{-10pt}}
\caption{\corrl{} Pattern Implementation in Context/Capability-Passing Style.}\label{polyjoins:fig:cartesiuswhereyield}
\end{figure}

%%% Local Variables:
%%% mode: latex
%%% TeX-master: "report"
%%% End:

\begin{figure}
  \centering
\begin{adjustbox}{valign=T,left,scale=.9}
\begin{lstlisting}[aboveskip=0pt,belowskip=-10pt]
let join: type s a b. (s, a) ctx -> s ext -> (a -> (s, b) pat) -> b shape repr
  = fun ctx extension pattern_body ->
      (* (effect Push$\commentcolor_i$: s$\commentcolor_i$ -> unit)$\commentcolor_{1\leq i \leq n}$: *)
      let slots: s ?Slots?.hlist = gen_slots_from ctx in
      (*  effect Yield: b -> unit and effect Fail: unit -> 'c: *)
      let yf: b yieldfail = gen_yieldfail () in
      (* instantiate and run bottom-right corner of Figure (*@\ref{polyjoins:fig:cartesius-overview}@*) *)
      let pstreams = parallel_bind ctx slots in
      let h$_\otimes$        = gen_default_handler slots yf pattern_body in
      let h_user   = extension ctx in
      let h_sys    = gen_sys_handler yf in
      ?Async?.spawn (h_sys |+| h$_\otimes$ |+| h_user) pstreams
\end{lstlisting}
\end{adjustbox}\myhack{\vspace{-10pt}}
 \caption{Implementation of \corrl{} Join Patterns.}\label{polyjoins:fig:cartesiusjoinimpl}
\end{figure}

%%% Local Variables:
%%% mode: latex
%%% TeX-master: "report"
%%% End:

\subsection{Extending PolyJoin}\label{polyjoins:sec:polyjoin-with-meta}

We extend core \polyjoin{} from \Cref{polyjoins:sec:solution} with metadata and contextual
extensions to provide a type-safe embedding of the \corrl{}
language. \Cref{polyjoins:fig:polyjoin-cartesius} defines the extended \polyjoin{} rules,
which we discuss in the following.

\subsubsection{Meta Data}\label{polyjoins:sec:meta-data}
The abstract type $\Type{Meta}$,
represents metadata carried by event values from event sources. \Eg, the occurrence times of events
(as in \corrl{}) or geolocation coordinates as in EventJava~\cite{Eugster:2009aa}. We expose meta
data for each variable in the join pattern body, by changing the previous shape translation judgment
accordingly (rules \inflabel{tz} and \inflabel{ts}). Metadata representations can be merged using
the binary operator $\sqcup$
(rule \inflabel{mmerge}). In the HOAS variable representation, end users can reuse OCaml's deep
pattern matching on function parameters to decompose the bound event variables into value and meta
datum, \eg, \Cref{polyjoins:fig:cartesius-combine-latest},
\Cref{polyjoins:lst:cartesius-combine-latest:deep}.

A tagless interpreter requires a policy for computing the metadata of joined events from the meta
data of input events. However, we do not expose merging explicitly in the pattern syntax. \Eg, the
pattern body in \Cref{polyjoins:fig:cartesius-combine-latest} does not mention the output event's
meta datum in the \lstinline|yield| form. This design reduces syntactic noise in the pattern and
avoids that users incorrectly merge the metadata for the resulting event, potentially violating
invariants of the underlying system.

While merging is implicit in the end user pattern syntax, it should be explicitly stated in
the syntax type signature, to obligate the tagless interpreter to provide an implementation
for merging. We modify the
pattern formation rules accordingly: The pattern formation judgment
$\formspatm{\vec{C}}{P}{A}$
now states that $P$
is a pattern yielding $A$
events and requires a metadata context derived from the shape $\vec{C}$.
By unification, the requirement $\vec{C}$
will be matched to the shape $\vec{A}$ of the pattern context in the $\inflabel{join}$ rule.

\subsubsection{Contextual Extensions}\label{polyjoins:sec:cont-restr}
In order to support restriction handlers in the pattern syntax, we introduce an
abstract syntax for contextual extensions $\formsext{\vec{A}}{X}$,
meaning that $X$
is an injectable extension, which depends on/can only be used in contexts of shape $\vec{A}$.
This is to ensure that restriction handlers as in
\Cref{polyjoins:fig:cartesius-combine-latest} cannot refer to non-existent positions in the context
of bindings.  Furthermore, we assume contextual extensions are monoids, having an empty extension
(rule \inflabel{eempty}) and a merge operation (rule \inflabel{emerge}), just as the restriction
handlers in \corrl{}.

\subsubsection{OCaml Representation}\label{polyjoins:sec:extended-ocaml-representation}
The extensions to \polyjoin{} presented here adapt straightforwardly to an OCaml module signature
encoding, in the same manner as before (\Cref{polyjoins:sec:ocaml-repr-polyj}).
Below, we show the key changes in the \lstinline|Symantics| signature:
\begin{lstlisting}[aboveskip=1pt,belowskip=1pt]
module type ?Symantics? = (* ... *)
  (* Contextual Extensions: *)
  val enil: unit -> 'a ext
  val (|++|): 'a ext -> 'a ext -> 'a ext
  (* Expressions and Patterns: *)
  val mmerge: meta repr -> meta repr -> meta repr
  val join: ('a,'b) ctx -> 'a ext -> ('b -> ('a,'c)) pat) -> 'c shape repr
end
\end{lstlisting}
The function \lstinline|mmerge| corresponds to the metadata merge operator $\sqcup$
in \Cref{polyjoins:fig:polyjoin-cartesius}. Due to space limitations, we
elide the full definition and leave it as an exercise to the reader.

\subsubsection{Predicates on Meta Data}\label{polyjoins:sec:patt-synt-design}
We sketch how the extended version of \polyjoin{} enables new predicates on metadata, which we can easily add in
the tagless final encoding. \Eg, in \corrl{}, events carry time intervals (pairs of time stamps) as
metadata: \lstinline|type meta = time * time|, where the \lstinline|time| is a numeric
type for time stamps (cf.~\cite{Bracevac:2018aa}), having a total order along with the following
operations:
\begin{lstlisting}[numbers=none,aboveskip=1pt,belowskip=1pt,xleftmargin=40pt]
val infty:   time repr  (* greatest element, positive infinity *)
val ninfty:  time repr     (* least element, negative infinity *)
val (%<=):   time repr -> time repr -> bool repr (* comparison *)
val span:    time repr -> time repr -> time repr  (* time span *)
val minutes: float -> time repr   (* representation of minutes *)
\end{lstlisting}

\noindent The \lstinline|within| constraint in our running fire alarm example
(\Cref{polyjoins:fig:fireexampleocaml}) is definable as ``derived syntax'' (\ie, function
abstraction) in the tagless DSL
\begin{lstlisting}[aboveskip=1pt,belowskip=1pt,numbers=none]
let within: meta repr -> meta repr -> time repr -> bool repr =
  fun m1 m2 mb -> (span (m1 $\sqcup$ m2)) %<= mb
\end{lstlisting}
where we compare the merged intervals (their least upper bound) against
the given time span.

\subsection{Embedding the Core of \corrl{} with PolyJoin}\label{polyjoins:sec:type-safe-polyv}

We provide a high-level outline of our \polyjoin{} tagless interpreter for \corrl{}, in the
multicore OCaml dialect. The interpreter is meta-circular, \ie, it interprets join patterns
as multicore OCaml code.

\subsubsection{Polyvariadic Effect Declarations}\label{polyjoins:sec:modul-gener-effects}
We obtain polyvariadic effect declarations by representing heterogeneous sequences of effect \emph{declarations}
in the language of \emph{types and expressions}.

First, we represent single effects as expressions/values.
We adopt a folklore encoding of
\emph{first-class effects}, which is already
utilized in Bra\v{c}evac et al.'s non-polyvariadic prototype.
\Cref{polyjoins:fig:slotsignature} shows the first-class effect encoding in OCaml.
A first-class effect is represented by a first-class module instance of the signature
\lstinline|?SLOT?|, carrying effect declarations. Their signature depends on the abstract
\lstinline|type t|, which represents the element type of an input event source. A slot module
carries an instance-specific \lstinline|&@Push&@| effect declaration, as well as
\lstinline|&@Set&@| and \lstinline|&@Get&@| effects for retrieving/updating a mailbox of local observations
(\Cref{polyjoins:sec:n-entangl-cont}).
Similarly, we encapsulate the
\lstinline|&@Yield&@| and \lstinline|&@Fail&@| effects (cf.~\Cref{polyjoins:fig:cartesius-overview}) in the
\lstinline|?YF?| module.

First-class effect instances are capability values that programmers
can pass around and manipulate.
\Eg, \Cref{polyjoins:lst:slotsignature:typealias0,polyjoins:lst:slotsignature:typealias1} of the
left definition defines the \lstinline|'a slot| capability type.  We use the \lstinline|with type| construct to
equate the type variable \lstinline|'a| with the abstract type \lstinline|t| in the respective
module. In this way, the abstract effect type becomes visible in the language of types.

Finally, we represent heterogeneous sequences of effect declarations
by heterogeneous lists of first-class effects values:
\begin{lstlisting}[numbers=none,aboveskip=1pt,belowskip=1pt]
module ?Slots? = ?HList?(struct type 'a t = 'a slot end)
\end{lstlisting}
Shape preservation (\Cref{polyjoins:sec:shape-preservation}) ensures that the effect types are
consistent with the input and variable types of the join computation. That is, if an $n$-way
join computation has shape \lstinline|'a|, then a value of type \lstinline|'a ?Slots?.hlist| carries
$n$ capabilities to push events, each matching the corresponding input source's type.
If an $n$-way
join computation has shape \lstinline|'a|, then a value of type \lstinline|'a ?Slots?.hlist| carries
$n$ capabilities to push events, each matching the corresponding input source's type.

\subsubsection{Heterogeneous Effect Handling}\label{polyjoins:sec:restriction-handlers-repr}
The first-class effect encoding enables heterogeneous handlers. We represent them by ordinary
handlers plus context, accepting the first-class \lstinline|slot| effects:
\begin{lstlisting}[aboveskip=1pt,belowskip=1pt,numbers=none]
type 'a handler  = (unit -> 'a) -> 'a
type 'ctx ext    = 'ctx ?Slots?.hlist -> unit handler
let (|++|) h1 h2 = fun ctx -> (h1 ctx) |+| (h2 ctx)
\end{lstlisting}
Values of type \lstinline|'ctx ext| may calculate a handler depending on these effects. Since we
model asynchronous processes that do not return, we let the handlers have the \lstinline|unit|
return type. For example, the function
\begin{lstlisting}[aboveskip=1pt,belowskip=1pt]
let print_pushes: type a. (int * (string * a)) ext = fun slots ->
  module ?IntS? = (val (head slots)) in
  module ?StrS? = (val (head (tail slots))) in
  fun action -> try action () with
    | effect (?StrS?.&@Push&@ s) k -> println(s); continue k ()
    | effect (?IntS?.&@Push&@ i) k -> println(int_to_str(i)); continue k ()
\end{lstlisting}
calculates a handler that prints integer- and string-valued push-notifications
to the console, by accessing the respective first-class effect instances
and handling their \lstinline|&@Push&@| effects. The point is that we can simultaneously handle
heterogeneous instantiations of these effects in a type-correct way within one handler.
\Ie, the \lstinline|print_pushes| handler handles both a string- and integer-valued instance
of \lstinline|&@Push&@|. The variants can be discerned by assigning the first-class effects
to a local module declaration (Lines 2-3), and then qualify via the module's name
the intended effect declaration (Lines 5-6).

\subsubsection{Context Polymorphism}\label{polyjoins:sec:context-polymorphism}
Moreover, the \lstinline|print_pushes| handler above is safely applicable in infinitely many contexts,
because it is parametric over all context shapes \lstinline|(int * (string * a))|, for all
\lstinline|a|. We call this trait \emph{context polymorphism}.
More generally, we can define polyvariadic handlers, by calculations over heterogeneous lists. \Eg,
\Cref{polyjoins:fig:polyhandler} shows the polyvariadic handler \lstinline|memory|.  This handler is
part of $h_\otimes$
in \Cref{polyjoins:fig:cartesius-overview} and handles the mailbox effects. That is, it maintains
$n$
distinct mailbox states in reference cells (\Cref{polyjoins:fig:polyhandler:cell}) and handles $n$
distinct \lstinline|&@Set&@| and \lstinline|&@Get&@| effects, by reading/writing the corresponding
mailbox.  Due to space limitations, we elide the definition of the \lstinline|poly_handler|
combinator. It is essentially mapping the supplied function
over the $n$
first-class effects and then composing the resulting handlers into a single handler.  The more
general takeaway from this example is that context polymorphism enables polyvariadic
components and their composition, \ie, an entire polyvariadic backend implementation
for the polyvariadic frontend of \polyjoin{}.

\subsubsection{Implicit Time Data in Patterns}\label{polyjoins:sec:repr-cont-depend}
We let the syntactic sort of patterns denote
functions types as one possible representation of context dependency, \ie,
\begin{lstlisting}[numbers=none,aboveskip=1pt,belowskip=1pt]
type ('c,'a) pat = 'c ?Meta?.hlist -> 'a yieldfail -> 'a repr * meta repr
\end{lstlisting}
accepting a heterogeneous list of time stamps, \ie,
\begin{lstlisting}[numbers=none,aboveskip=1pt,belowskip=1pt]
module ?Meta? = ?HList?(struct type 'a t = meta repr end)
\end{lstlisting}
and a capability to yield and fail. The end result is a pair of output event together with its meta
datum, which is derived from the metadata of the input events.

Metadata and capabilities must be explicitly threaded by the tagless interpreter of the pattern
syntax. Their essential uses occur in the implementations of \lstinline|where| and \lstinline|yield|
(\Cref{polyjoins:fig:cartesiuswhereyield}).  In the case of \lstinline|where|, if all the
constraints are satisfied in \lstinline|cond|, we continue with the body, passing down the
context. Otherwise we cancel the current pattern match attempt by invoking the \lstinline|&@Fail&@|
effect from the given capability \lstinline|yf|. The failure invocation is abstracted in the
function \lstinline|fail_with yf|.  In the case of \lstinline|yield|, we return the given result and
merge the implicit metadata of the input with \lstinline|merge_all: 'a ?Meta?.hlist -> meta|.

The point is that the function-based pattern type interpretation forces a tagless
interpreter to supply a value for implicitly merging metadata (\eg, extracted from the $n$
tuple of input events) before a join pattern can produce an event.

\subsubsection{Join Pattern Implementation}\label{polyjoins:sec:putting-it-together}
\Cref{polyjoins:fig:cartesiusjoinimpl} shows the implementation of the join
pattern form.
The code is a direct translation of the diagram in \Cref{polyjoins:fig:cartesius-overview}. First,
it allocates first-class effect instances and then continues setting up the concurrent join
computation. The last line corresponds to executing the bottom-right corner of
\Cref{polyjoins:fig:cartesius-overview} in an asynchronous thread.  Due to space limitations, we
omit the implementations of the involved combinators.
The point is that the join signature
in \polyjoin{} indeed supports non-sequential event binding, since this is a concurrent join
computation reacting to events as they come.

And it is statically type-safe!
The type signature certifies that (1) \lstinline|join| is polyvariadic, (2) that supplied
restriction handlers cannot refer to non-existing input sources and (3) the implementation
discharges the implicit metadata requirement by the HOAS pattern abstraction
\lstinline|(a -> (s,b)$\,$pat)|.
To see the latter point, we first inline the type definition for patterns into the function type, obtaining
\begin{lstlisting}[numbers=none,aboveskip=1pt,belowskip=1pt]
a -> s ?Meta?.hlist -> b yieldfail -> b repr * meta repr      $(\dagger)$
\end{lstlisting}
as the type of the pattern body. By a simple induction over the derivation of the
context representation \lstinline|(s,a)$\,$ctx|, we can establish that
(with some generous typographic simplification)
\begin{lstlisting}[numbers=none,aboveskip=1pt,belowskip=1pt]
a = (a$_1$ repr * meta repr) * ... * (a$_n$ repr * meta repr)
s = a$_1$ * ... * a$_n$
\end{lstlisting}
where $n$
is the arity of the join.  Therefore, by definition of \lstinline|?Meta?.hlist|, we have that the
parameter type \lstinline|s ?Meta?.hlist| is an $n$ tuple of metadata.
Importantly, the \emph{implementation} and not the end user supplies the metadata to be merged.

By parametricity (cf. \cite{Wadler:1989ab}), the signature of \lstinline|join| and
$(\dagger)$
even determines how the \corrl{} implementation functions. It extracts $n$-tuples
and metadata from the event sources and tests them against the pattern, potentially\footnote{It may
  throw a failure via the supplied \lstinline|yieldfail| capability.}  resulting in an output
\lstinline|b repr * meta repr|. This is the only way to obtain \lstinline|b|-typed events, which may
be communicated over the output event source \lstinline|b shape repr|.  However, the types are too
weak to enforce that the implementation is always productive, \ie, it might be inert or diverging.
More sophisticated type systems are necessary to enforce productivity, \eg, where types represent
theorems in linear temporal logic (LTL)~\cite{Cave:2014aa}.

\subsubsection{Solving the Focusing Continuations Problem}\label{polyjoins:sec:cart-prod-revis}
Defining a polyvariadic version of \corrl{}' callback logic (Sections
\ref{polyjoins:sec:n-entangl-cont}, \ref{polyjoins:sec:cartesius-overview:summary} and
\Cref{polyjoins:fig:focusing}) was the most challenging part of the implementation.  Fortunately,
the tools we have so far enable a simple solution (\Cref{polyjoins:fig:focusing-solution}), which does not
require type-level zippers. We exploit that effect handling is a form of dynamic overloading.

The combinator \lstinline|callback| represents \emph{all} the callbacks $(\kappa_i)_{1\leq i\leq n}$
from \Cref{polyjoins:fig:focusing}. It is a polyvariadic handler for the \lstinline|&@Push&@$_i$|
effects, so that the effect handling clause of the $i$-th
effect implements callback $\kappa_i$
(Lines~\ref{polyjoins:lst:callbacks:start}-\ref{polyjoins:lst:callbacks:end}). For an event
notification \lstinline|x|, we first retrieve the mailboxes of the join computation, replacing the
$i$-th
position by \lstinline|x|, via the \lstinline|focus| combinator. Then, we compute the cartesian
product over these mailboxes, passing all the $n$-tuples
to a \lstinline|consumer| function. This parameter represents the part of the system that tests
tuples against the join pattern, yielding or failing.

The \lstinline|focus| combinator is at the core of the solution, exploiting a synergy with effect
handlers, to obtain a type-safe focus and replacement into the heterogeneous list of mailboxes
(modeled by the type \lstinline|a ?Mail?.hlist|).
Lines \ref{polyjoins:lst:focus:start}-\ref{polyjoins:lst:focus:end} of \lstinline|focus| codify the
approach. We retrieve all mailboxes by invoking all of the $n$
\lstinline|&@Get&@| effects polyvariadically (function \lstinline|get_all|). To focus into the given
position, we \emph{locally} handle the $i$-th
\lstinline|&@Get&@| effect and \emph{override} its meaning, \ie, we answer the effect invocation for
the focused position by passing the supplied event \lstinline|x| in a mailbox.  For the non-focused positions,
\lstinline|focus| does not handle the \lstinline|&@Get&@| invocations, instead delegating the
handling to the calling context. We assume that the \lstinline|memory| handler
(\Cref{polyjoins:fig:polyhandler}) is in the context, to define the default semantics.

Finally,
\Cref{polyjoins:fig:stacklayout} summarizes our focusing solution graphically. Each box represents a
stack frame of the dynamic execution context, with the bottom-most being currently executed.
We effectively implement \emph{dynamic binding}, using the deep binding strategy~\cite{Moreau:1998aa},
in terms of effects (dynamic variables) and stacking handlers (dynamic variable bindings).

\begin{figure}
  \centering
  \begin{align*}
  \kappa_{i} :\ & A_i\to M[A_1\times\cdots\times A_n] \\
  \kappa_{i} =\ & \lambda \textcolor{ACMRed}{x}:A_i. m_1\otimes\cdots \otimes m_{i-1}\otimes\textcolor{ACMRed}{\anglep{x}}\otimes m_{i+1}\otimes\cdots \otimes m_n\quad 1\leq i \leq n\\
 (\_ \otimes \_) :\ & \forall \alpha\, \beta.M[\alpha] \to M[\beta] \to M[\alpha \times \beta]
\end{align*}\myhack{\vspace{-20pt}}
\caption{The Focusing Continuations Problem.}\label{polyjoins:fig:focusing}
\end{figure}

%%% Local Variables:
%%% mode: latex
%%% TeX-master: "report"
%%% End:

\begin{figure}
  \centering
  \begin{adjustbox}{valign=T,left,scale=.9}
\begin{lstlisting}[aboveskip=0pt,belowskip=0pt]
(* Callback logic *)
let callback slots consumer = poly_handler slots (fun (si: (module ?SLOT?)) ->
  let module ?Si? = (val si) in
  fun action -> try action () with
  | effect (?Si?.&@Push&@ $\textcolor{ACMRed}{\mathtt{x}}$) k -> (*@\label{polyjoins:lst:callbacks:start}@*)
      (* $\commentcolor m_1,\ldots, m_{i-1},\textcolor{ACMRed}{\anglep{\mathtt{x}}}, m_{i+1},\ldots , m_n$ *)
      let mboxes = focus slots s $\textcolor{ACMRed}{\mathtt{x}}$ in
      (* $\commentcolor m_1\otimes\cdots\otimes m_{i-1}\otimes\textcolor{ACMRed}{\anglep{\mathtt{x}}}\otimes m_{i+1}\otimes\cdots \otimes m_n$ *)
      crossproduct mboxes consumer; continue k () (*@\label{polyjoins:lst:callbacks:end}@*)
\end{lstlisting}
  \end{adjustbox}
 \begin{adjustbox}{valign=T,left,scale=.9}
\begin{lstlisting}[aboveskip=3pt,belowskip=0pt]
(* Focusing into $\commentcolor m_1,\ldots, m_n$ *)
let focus type a b. a ?Slots?.hlist -> b slot -> b ev -> a ?Mail?.hlist =
  fun slots si x ->
    let module ?Si? = (val si) in
    (* Replace i-th position by handling *)
    try get_all slots () with (*@\label{polyjoins:lst:focus:start}@*)
    | effect (?Si?.&@Get&@ ()) k -> continue k (mailbox x) (*@\label{polyjoins:lst:focus:end}@*)
\end{lstlisting}
 \end{adjustbox}
 \begin{adjustbox}{valign=T,left,scale=.9}
\begin{lstlisting}[aboveskip=3pt,belowskip=-10pt]
(* Bulk retrieval of all mailboxes $\commentcolor m_1,\ldots, m_n$ *)
let get_all: type a. a ?Slots?.hlist -> unit -> a ?Mail?.hlist = fun slots () ->
  let module ?M? = ?HMap?(?Slots?)(?Mail?) in
    ?M?.map { ?M?.f = fun (s': (module ?SLOT?)) ->
      let module ?S'? = (val s') in perform (?S'?.&@Get&@ ()) } slots
\end{lstlisting}
\end{adjustbox}\myhack{\vspace{-10pt}}
\caption{Multicore OCaml Solution to the Focusing Continuations Problem.}\label{polyjoins:fig:focusing-solution}
\end{figure}

%%% Local Variables:
%%% mode: latex
%%% TeX-master: "report"
%%% End:

\tikzstyle{boxtitle} = [draw=black,rectangle, font=\footnotesize\mdseries,inner sep=1.5pt]
\tikzstyle{stackframe} = [rectangle,box,draw=white]
\tikzstyle{handlerframe} = [very thick,draw=black,rectangle,box]
\begin{figure}
  \centering
  \begin{adjustbox}{valign=T,scale=.7}
\begin{tikzpicture}[box/.style={draw,minimum width=8cm,align=center},node distance=0pt]
    \node[stackframe] (Memory) {\\[3pt]\lstinline|(effect ?S?$_j$.&@Get&@ () k -> continue k !mbox$_j$)$_{1\leq j\leq n}$|};
    \node[box,draw=white,below= of Memory] (Misc) {$\vdots$};
    \node[stackframe,below= of Misc] (Focus)  {\\[3pt]\lstinline|$\phantom{(}$effect $\textcolor{ACMRed}{\mathtt{S}_i.\mathtt{Get}\ \texttt{()}}$ k -> continue k (mailbox $\textcolor{ACMRed}{\mathtt{x}}$)|};
    \node[stackframe,below= of Focus] (Getall) {\\[3pt]\lstinline|$\langle$?S?$_1$.&@Get&@ (),$\ldots$,$\textcolor{ACMRed}{\mathtt{S}_i.\mathtt{Get}\ \texttt{()}}$,$\ldots$,?S?$_n$.&@Get&@ ()$\rangle$|};
    \draw (Memory.south west) -- (Memory.south east);
    \draw (Misc.south west) -- (Misc.south east);
    \draw (Focus.south west) -- (Focus.south east);
    \node[
      draw,
      inner sep=0pt,
      fit={(Memory) (Misc) (Focus) (Getall)},
    ] (fit) {};
    % box labels:
    \node[boxtitle, below right] at (Memory.north west) {\lstinline|memory|};
    \node[boxtitle, below right] at (Focus.north west)  {\lstinline|focus$_i$|};
    \node[boxtitle, below right] at (Getall.north west) {\lstinline|get_all|};
%    \node[below right, inner sep=0pt] at (fit.north west) {below right};
\end{tikzpicture}
\end{adjustbox}\myhack{\vspace{-8pt}}
\caption{Illustration of Dynamic Execution Context during Focusing.}\label{polyjoins:fig:stacklayout}
\end{figure}
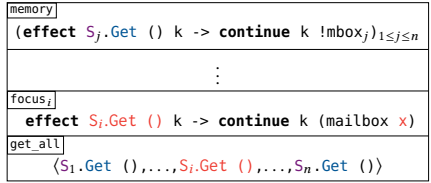
%%% Local Variables:
%%% mode: latex
%%% TeX-master: "report"
%%% End:

\subsection{Safe, Reusable and Modular Restriction Handlers}\label{polyjoins:sec:extens-writ-restr}

\begin{figure}
\begin{minipage}[t]{.5\linewidth}%
  \begin{adjustbox}{left,valign=T,scale=.9,margin=19pt 0pt 0pt 0pt}
\begin{lstlisting}[aboveskip=0pt,belowskip=-10pt]
(* (*@$\commentcolor\ctxproj{n}{A}{\vec{B}}$@*) (Figure (*@\ref{polyjoins:fig:polyjoin-cartesius}@*)) *)
type (_,_) _ptr =
  | Pz: ('a, 'a * 'b) _ptr
  | Ps: ('a, 'b) _ptr
           -> ('a, 'c * 'b) _ptr
type ('a,'b) ptr =
  unit -> ('a,'b) _ptr
\end{lstlisting}%
  \end{adjustbox}
\end{minipage}%
\begin{minipage}[t]{.5\linewidth}%
  \begin{adjustbox}{left,valign=T,scale=.9}
\begin{lstlisting}[aboveskip=0pt,belowskip=-10pt]
(* (*@$\commentcolor\ctxmproj{\vec{n}}{\vec{A}}{\vec{B}}$@*) (Figure (*@\ref{polyjoins:fig:polyjoin-cartesius}@*)) *)
type (_,'a) _mptr =
  | Mz: (unit, 'a) _mptr
  | Ms: (('c, 'b) _ptr * ('a, 'b) _mptr)
            -> ('c * 'a, 'b) _mptr
type ('a,'b) mptr =
  unit -> ('a,'b) _mptr
let mz ()      = Mz
let ms p ps () = Mn (p (), ps ())
\end{lstlisting}%
  \end{adjustbox}
\end{minipage}\myhack{\vspace{-10pt}}
\caption{GADTs for Type-safe Pointers and Sets of Pointers.}\label{polyjoins:fig:pointers}
\end{figure}

%%% Local Variables:
%%% mode: latex
%%% TeX-master: "report"
%%% End:

\begin{figure}
  \centering
\begin{adjustbox}{valign=T,left,margin=0pt 0pt 3pt 0pt,scale=.9}
\begin{lstlisting}[belowskip=0pt,aboveskip=0pt,escapechar=!]
let p0 () = Pz       (* (!\textcolor{ACMRed}{'a}!, !\textcolor{ACMRed}{'a}! * 'b) ptr *)
let p1 () = Ps Pz    (* (!\textcolor{ACMRed}{'a}!, 'b * (!\textcolor{ACMRed}{'a}! * 'c)) ptr *)
let p2 () = Ps Ps Pz (* (!\textcolor{ACMRed}{'a}!, 'b * ('c * (!\textcolor{ACMRed}{'a}! * 'd))) ptr *)
let ps () =          (* (!\textcolor{ACMRed}{'a}! * (!\textcolor{ACMRed}{'b}! * unit), !\textcolor{ACMRed}{'b}! * ('c * (!\textcolor{ACMRed}{'a}! * 'd))) mptr *)
  (ms p2 @@ ms p0 @@ mz) ()
\end{lstlisting}
\end{adjustbox}\myhack{\vspace{-10pt}}
\caption{Example: Type-safe Pointers and Sets of Pointers in OCaml.}\label{polyjoins:fig:type-safe-pointers}
\end{figure}

%%% Local Variables:
%%% mode: latex
%%% TeX-master: "report"
%%% End:

\begin{figure}
  \begin{adjustbox}{valign=T,scale=.9,left,margin=19pt 0pt 0pt 0pt}
\begin{lstlisting}[aboveskip=0pt,belowskip=-15pt]
module ?HPtr?(?H?: ?Hl?) = struct
  let rec proj: type a ctx. (a, ctx) _ptr -> ctx ?H?.hlist -> a ?H?.el =
    fun n hlist -> match n, hlist with
      | Pz,   ?H?.S (hd, _) -> hd
      | Ps n, ?H?.S (_, tl) -> proj n tl

  let rec mproj: type xs ctx. (xs, ctx) _mptr -> ctx ?H?.hlist -> xs ?H?.hlist =
    fun mptr hlist -> match mptr with
      | Mz         -> ?H?.nil
      | Ms (i, ps) -> ?H?.cons (proj i hlist) (mproj ps hlist)
end
\end{lstlisting}
  \end{adjustbox}\myhack{\vspace{-20pt}}
\caption{Type-safe Element Access.}\label{polyjoins:fig:projection}
\end{figure}

%%% Local Variables:
%%% mode: latex
%%% TeX-master: "report"
%%% End:

\begin{figure}
  \centering
\begin{adjustbox}{valign=T,left,scale=.9}
\begin{lstlisting}
let most_recently: type ctx i. (i, ctx) ptr -> ctx ext = fun ptr slots ->
  (* Type-safe pointers and projection on first-class effects *)
  let module ?SPtr? = ?HPtr?(?Slots?) in (*@\label{polyjoins:lst:most-recently:access:start}@*)
  (* Project the given first-class effect declaration *)
  let module ?Si? = (val (?SPtr?.proj (ptr ()) slots)) in (*@\label{polyjoins:lst:most-recently:access:end}@*)
  fun action -> try action () with (*@\label{polyjoins:lst:most-recently:body:start}@*)
    | effect (?Si?.&@Push&@ x) k ->
        (* Truncate the mailbox to the most recent value *)
        perform ?Si?.&@Set&@ (mailbox x); continue k (perform ?Si?.&@Push&@ x) (*@\label{polyjoins:lst:most-recently:body:end}@*)
\end{lstlisting}
\end{adjustbox}\myhack{\vspace{-10pt}}
\caption{Example: Position-polymorphic Restriction Handler (\lstinline|most_recently|).}\label{polyjoins:fig:most-recently}
\end{figure}

%%% Local Variables:
%%% mode: latex
%%% TeX-master: "report"
%%% End:

Here, we solve the challenge of writing restriction handlers generically
(\Cref{polyjoins:sec:cartesius-overview:summary}).  Following the design of \citet{Bracevac:2018aa},
it should be enough to know about the \lstinline|&@Push&@|, \lstinline|&@Get&@| and
\lstinline|&@Set&@| effect interfaces in order to write restrictions. These interfaces are a natural
abstraction barrier, to not burden programmers with the details of the underlying join
implementation.
We show a simple solution for programming restriction
handlers that is both \emph{context-polymorphic} and \emph{position-polymorphic}, so that one
definition is compatible with join instances of different arity and we have fine-grained control
where to apply restrictions.

\subsubsection{Type-Safe Pointers/De Bruijn Indices}\label{polyjoins:sec:type-safe-pointers}
Recall that we apply restriction handlers to specific positions via De Bruijn indices (\eg,
\Cref{polyjoins:fig:cartesius-combine-latest},
\Cref{polyjoins:lst:cartesius-combine-latest:mostrecently}).  Formally, a De Bruijn index
corresponds to a derivation of the shape projection judgment $\ctxproj{n}{A}{\vec{B}}$
having rules \inflabel{pz} and \inflabel{pz} (\Cref{polyjoins:fig:polyjoin-cartesius}).  The
judgment asserts: the index number $n$
proves that type $A$
is contained in the heterogeneous sequence $\vec{B}$.
The rules of shape projection straightforwardly translate to a binary GADT \lstinline|_ptr|
(\Cref{polyjoins:fig:pointers}, left), which is well-known by functional programmers.  This leads to
the type \lstinline|ptr| of \emph{type-safe pointers} into heterogeneous contexts, which communicate
\emph{context requirements} at the type-level.  We show examples of type-safe pointers along with
their types in \Cref{polyjoins:fig:type-safe-pointers}.  The pointer type relates to the formal
system in \Cref{polyjoins:fig:polyjoin-cartesius} in that we let the type of a pointer value $n$
represent a polymorphic \emph{judgment scheme}, from which all derivable ground judgments
$\ctxproj{n}{A}{\vec{B}}$
can be instantiated. We can interpret the scheme as a \emph{context requirement}.  \Eg, the
polymorphic type\footnote{The distinction between \lstinline|_ptr| and \lstinline|ptr| is necessary,
  due to the value restriction in ML/OCaml~\cite{Wright:1995aa}. The pointers must be functions in
  order to fully generalize the type parameters.}  of the pointer \lstinline|p1| certifies that it
points at the second element \lstinline|'a$\,$| (highlighted in red) of contexts \emph{having at least
  two elements}. This is due to the context shape \lstinline|'b * ('a * 'c)| being
polymorphic in the tail \lstinline|'c|.

By induction over the shape projection derivation, we can leverage these type-level
assertions to define a type-safe projection function \lstinline|proj| on heterogeneous
lists, where projecting the $i$-th
element always succeeds (\Cref{polyjoins:fig:projection}). \Ie, if \lstinline|(i,a)$\,\_$ptr|
holds, then any \lstinline|hlist| of shape \lstinline|a| contains an \lstinline|i|-typed element,
which can be retrieved. The definition we show is contained in a functor \lstinline|?HPtr?|, to make
the projection function parametric over all the ``uniformly heterogeneous'' lists
(\Cref{polyjoins:sec:constr-heter}).

\subsubsection{Sets of Type-Safe Pointers}\label{polyjoins:sec:sets-type-safe}
On top of shape projection, we define the shape multiprojection judgment
$\ctxmproj{\vec{n}}{\vec{A}}{\vec{B}}$
(\Cref{polyjoins:fig:polyjoin-cartesius}), which is a straightforward extension.  It reads: the
sequence $\vec{n}$
of De Bruijn indices points at multiple positions having types $\vec{A}$
within a heterogeneous sequence $\vec{B}$.
The corresponding OCaml GADT \lstinline|_mptr| is defined in \Cref{polyjoins:fig:pointers} and
enables sets of type-safe pointers (type \lstinline|mptr|). Similarly to the type-safe pointers, the
sets describe polymorphic judgment schemes for context requirements.  For example, the set of
pointers \lstinline|ps| in \Cref{polyjoins:fig:type-safe-pointers} points into the third and first
positions, and requires a context of at least three elements.  Finally, we define a type-safe
multiprojection function \lstinline|mproj| (\Cref{polyjoins:fig:projection}), which guarantees that
projecting the pointed-at positions in the pointer set yields the heterogeneous list of projected
values.

\subsubsection{Position Polymorphism for Restriction Handlers}\label{polyjoins:sec:position-polymorphism}
Being parametric over type-safe pointers/sets enables library writers to impose more refined
constraints on context-polymorphic functions (cf.~\Cref{polyjoins:sec:context-polymorphism}).  We
call this refinement \emph{position polymorphism}, and it enables generic restriction handlers for
\corrl{}. For instance, we define the \lstinline|most_recently| restriction in
\Cref{polyjoins:fig:most-recently}.  This restriction changes the semantics of the $i$-th
pattern variable, by truncating past event observations.  \Ie, it handles the $i$-th
\lstinline|&@Push&@| effect of the join and overwrites the contents of the $i$-th
mailbox with just the current event notification
(Lines~\ref{polyjoins:lst:most-recently:body:start}-\ref{polyjoins:lst:most-recently:body:end}).
Accessing the required $i$-th
first-class effect instance is just a matter of projecting it from the available instances, using
the given pointer
(Lines~\ref{polyjoins:lst:most-recently:access:start}-\ref{polyjoins:lst:most-recently:access:end}).
The type-safe pointers guarantee that access always succeeds. Analogously, restriction handlers on
sets, \eg, \lstinline|aligning| (cf.~\Cref{polyjoins:sec:restriction-handlers}), are
functions parametric over type-safe pointer sets
\begin{lstlisting}[numbers=none,aboveskip=1pt,belowskip=1pt]
aligning: type xs ctx. (xs,ctx) mptr -> ctx ext
\end{lstlisting}
which use the multiprojection function to focus on the given positions \lstinline|xs| in \lstinline|ctx|.
Due to space limitations, we elide the definition of this handler and refer to \cite{Bracevac:2018aa}.

\subsubsection{Summary}\label{polyjoins:sec:demand-meets-supply}
Our context- and position-polymorphic function definitions give important static guarantees for both
system programmers and end users, checked and enforced by the OCaml compiler. First, recall that we
keep track of the join shape $\vec{A}$
in the context formation judgment (\Cref{polyjoins:fig:polyjoin-core}).  The type variable
\lstinline|'a| in the abstract type \lstinline|('a,'b)$\,$ctx|
is the OCaml equivalent of the shape. It serves as a type-level index and ``glue'' that binds
polyvariadic backend components and restriction handlers together. Components that share the same
shape \lstinline|'a| in their signature are composable.

Context polymorphism (\Cref{polyjoins:sec:context-polymorphism}) ensures that we can write
polyvariadic components that compose with the backend implementation of any join instance of
arbitrary shape. The type-level shape \lstinline|'a| is usually accompanied by value-level content,
which we think of as a polyvariadic interface for embedding a component into the backend implementation.
In the case of \corrl{}, the accompanying content are the first-class effect
declarations of type \lstinline|'a ?Slots?.hlist| (\Cref{polyjoins:sec:modul-gener-effects}), from
which we calculate effect handlers (\eg, \Cref{polyjoins:fig:polyhandler}).

Position polymorphism ensures that we can write polyvariadic components that only need
access to a specific part of the join shape \lstinline|'a|.  By construction, the access ``cannot go
wrong'', because we track usage context requirements in type-safe pointers and sets. We simply
demand in the signature of position-polymorphic definitions that the requirement equals the abstract
join shape against which we write the implementation. \Eg, \lstinline|('i,'a)$\,$ptr -> 'a ext| for
single positions (\Cref{polyjoins:fig:most-recently}), respectively
\lstinline|('xs,'a)$\,$mptr -> 'a ext| for sets of positions.

Furthermore, type-safe pointers and sets prevent end users from applying restrictions to wrong
positions. \Ie, referenced positions always are within bounds and have types that are compatible
with the requirements of a restriction.  Again, we ensure this by matching the join shapes. \Ie, rule \inflabel{join}
(\Cref{polyjoins:fig:polyjoin-cartesius}) matches the join shape demanded by the given extensions
to the join shape $\vec{A}$ supplied by the join pattern syntax.

%%% Local Variables:
%%% mode: latex
%%% TeX-master: "report"
%%% End:

\section{Evaluation and Discussion}\label{polyjoins:sec:evaluation}

\begin{table}[t]
\caption{Comparison between \polyjoin{} Version and Prototype of \corrl{}.}\label{polyjoins:tab:comparison}\myhack{\vspace{-8pt}}
\begin{minipage}{1.0\linewidth} %this gives footnotes a local scope
\centering
\begin{adjustbox}{max width=1.0\textwidth,scale=.8}
\begin{tabular}[t]{l@{\hskip 2em}c@{\ \ }c@{\ \ }}
  \hline\hline
                                 & \polyjoin{} & \textsc{Prototype} \\
  \hline
  \multicolumn{3}{l}{\textsc{Features}} \\
  {\quad Heterogeneity}          & \OK & \OK \\
  {\quad Arity-Genericity}       & \OK & \NAY\\
  {\quad Context Polymorphism}   & \OK & \NAY \\
  {\quad Position Polymorphism}  & \OK & \NAY \\
  {\quad Declarative Pattern Syntax}     & \OK & \NAY \\[.8ex]
  \multicolumn{3}{l}{\textsc{Implementation Code Size}\footnote{$n = $ maximum arity in use.}}\\
  {\quad Core (Fig.~\ref{polyjoins:fig:cartesius-overview})}                   & $O(1)$ & $O(n^2)$ \\
  {\quad Restriction Handlers (Over $c$ Positions)}   & $O(1)$ & $O(n^{c+1})$\\
  {\quad Restriction Handlers (Over Position Sets)}   & $O(1)$ & $O(2^n)$\\[.8ex]
  \textsc{Join Instance Size}\footnote{$n = $ number of inputs.} & $O(n)$ & $O(n)$ \\[.8ex]
  \multicolumn{3}{l}{\textsc{Extensibility Dimensions}}\\
  {\quad Syntax}                 & \OK & \NAY \\
  {\quad Restriction Handlers}   & \OK & \SOSO\footnote{Requires separate copy per supported arity.} \\
  {\quad Computational Effects}   & \OK & \OK \\
  \hline\hline
\end{tabular}
\end{adjustbox}
\end{minipage}
%\vspace{-2\baselineskip}
\end{table}

%%% Local Variables:
%%% mode: latex
%%% TeX-master: "report"
%%% End:

\Cref{polyjoins:tab:comparison} compares our \polyjoin{}-based implementation
of \corrl{} against the original prototype by \citet{Bracevac:2018aa}.
Our version has significant advantages over the original prototype. We discuss
them below.

\subsection{Features}\label{polyjoins:sec:evaluation:features}

In terms of features, the \polyjoin{} version of \corrl{} is fully polyvariadic, \ie, it
supports heterogeneous sources and any finite join arity. The original prototype satisfies ``one
half'' of polyvariadicity, \ie, its backend does not support arity abstraction and offers only a
handful of hard-coded arities. Each supported arity entails a separate copy of the backend code that
has to be separately maintained, with little to no code sharing.  Similarly, absence of context and
position polymorphism (\Cref{polyjoins:sec:demand-meets-supply}) greatly increases programming
effort for both the backend and restriction handlers (we elaborate the issue below). Moreover, the
prototype has no declarative frontend syntax for join patterns, so that end users require detailed
knowledge of backend components and their composition to specify joins.

\subsection{Asymptotic Code Size}\label{polyjoins:sec:evaluation:code-size}

Hard-coded arities, lack of arity abstraction and lack of context/position polymorphism lead to
significant code duplication and severely undermine the composability and extensibility of the
system. This is at odds with the goal to create an extensible and general event correlation system
as originally envisioned.  The issues become apparent when considering the \emph{code size} of the
respective implementations, reflecting programming effort.

In the prototype, the maximum supported
arity $n$
must be statically provisioned and accordingly, at most $n$
copies of the backend, one for each $i\in\Set{1,\ldots, n}$
have to be either manually programmed or pre-generated. If a client intends to specify
a join pattern that exceeds the current maximum arity, then a new copy of the backend
must be generated beforehand. In the worst case, the prototype backend has
$O(\Sigma_{i=1}^n i) = O(n^2)$
size. This is because each arity $i$
leads to the dependency on $i$
first-class effect declarations and contains effect handlers with $i$
cases for these effects (\eg, the \lstinline|memory| handler in \Cref{polyjoins:fig:polyhandler}).

Similar reasons apply in the case of restriction handlers, where we distinguish between restrictions
parametric in a constant number $c$
of positions (\eg, \lstinline|most_recently| in \Cref{polyjoins:fig:most-recently} has $c = 1$
position parameters) and restrictions parametric in position sets (\eg, \lstinline|aligning|
restriction \Cref{polyjoins:sec:restriction-handlers}).  In the prototype, a separate handler
definition must be written for \emph{each arity} and \emph{each combination} of positions,
respectively \emph{each subset} of $\Set{1,\ldots, n}$.
The code sizes become $O(n^{c+1})$
and $O(2^n)$,
accordingly.  Context and position polymorphism in \polyjoin{} reduce the programming effort to a
single definition and thus have constant code size. Clients can specify join patterns of arbitrary
arity and automatically, a suitable instance of the backend is calculated on demand.

In both versions, the size of a join computation at runtime is linear in the number $n$
of joined input sources, because a join computation allocates $n$
first-class effect instances. However, this is not a predictor of the overall runtime behavior,
\eg, computational cost and memory requirements during a join's execution. These measures are
highly dependent on the semantic variant specified by users. We refer to the measurements in
\cite{Bracevac:2018aa} for the runtime behavior of example join variants.

\subsection{Extensibility Dimensions}\label{polyjoins:sec:evaluation:extensiblity-dimensions}

The \polyjoin{} version of \corrl{} features an extensible pattern language, because of the
reliance on the tagless final approach (\Cref{polyjoins:sec:solution}).  New forms of syntax can be
easily added to the frontend language.

Restriction handlers are an orthogonal way to extend the system. They are sub-components that can be
separately developed, because only knowledge about an effect interface is required, represented by
the first-class \lstinline|?SLOT?| effects (\Cref{polyjoins:fig:slotsignature}).  However, the table
entry for the prototype is only half-checked (\SOSO). While new restriction handlers are
programmable, they are hardcoded against a specific join arity, due to the lack of context
polymorphism.

Finally, since both versions are designed around algebraic effects and effect handlers,
it is possible to induce new kinds of effects via restriction handlers and in the
\polyjoin{} case additionally in the implementations of new syntax forms.

\subsection{Discussion}\label{polyjoins:sec:discussion}

\myparagraph{Lessons Learned}
The \corrl{} case study heavily focuses on practical polyvariadic programming with
first-class effects and handlers. While this is already a significant contribution for
the algebraic effects community, it is also of value for the design and implementation of
asynchrony and concurrency languages as well as typed embeddings of process calculi.

(1) the techniques demonstrated are of general utility for implementing polyvariadic backends
against polyvariadic frontends: abstract type-level context shapes, context and position polymorphism
as well as heterogeneous lists for safely programming and composing polyvariadic sub-components.

(2) \polyjoin{} is a flexible and extensible design methodology that captures the essence
of join and synchronization pattern frontend syntaxes.

(3) polyvariadicity ``naturally'' occurs in concurrency and asynchrony systems, since
communication endpoints partaking in exchanges and synchronization are usually of heterogeneous
type and arbitrary in number. The interface of \Cref{polyjoins:def:joins} on which
we base \polyjoin{} is adequate to set up their synchronization logic, enabling
low-latency reactions and eliminating unnecessary blocking.

(4) the focusing problem we successfully solve in \Cref{polyjoins:sec:cart-prod-revis} is a manifestation
of external choice, which is an important aspect of concurrency.
While we implemented our solution
in terms of effects and handlers, it could be achieved with other dynamic binding
approaches, \eg, \cite{Kiselyov:2014aa}.

\myparagraph{Soundness of PolyJoin} The tagless final approach we build on ensures soundness of
patterns relative to the soundness of the host language's type system. The approach guarantees
\emph{by construction} that "well-typed event patterns (in the pattern DSL type system) cannot go
wrong". This is ensured, because we describe and check the DSL's type system via the host language's
type system. If the presented DSL variants and tagless interpreters were unsound, then they would be
a counterexample to the soundness of the host language. We effectively obtain a powerful safety-net
for developing variants of join pattern DSLs. If a programmer defines an unsound DSL, then it will
manifest itself at some point in the development, for example: (1) the syntax forms and DSL types
are ill-defined, which leads to the OCaml type checker rejecting the \lstinline|?Symantics?| module
signature, \ie, no tagless interpreter is implementable in the first place. (2) the
\lstinline|?Symantics?| signature is accepted, but then no useful programs could be formulated in
the functor representation of expressions.

\myparagraph{Abstraction Overhead}
Context access could be computed statically but is dynamic.
We believe that this is a negligible initial overhead. However
Multi-stage programming could help eliminating and optimizing overhead
from hlists and pointers.
\cite{Kiselyov:2014aa}
\cite{Rompf:2010aa}
But no support for effect handlers.

\myparagraph{Supported Systems} In general, any system is embeddable with \polyjoin{} for which a
suitable tagless interpreter can be written for the module signature in
\Cref{polyjoins:fig:patternsymantics}, respectively Section
4.2.3/Appendix A.1, if the system supports metadata in patterns.  These already permit a wide range
of backends.  First, any system that already has a LINQ-based frontend or is based on a monadic
embedding can be plugged into \polyjoin{} using the construction we show in
\Cref{polyjoins:fig:monadic-cartesian}.  Second, general, metadata-based event correlation backends
can be plugged into \polyjoin{}, which we have exemplified with \corrl{} in
\Cref{polyjoins:sec:case-study:-cart}.  Importantly, we support more than only library-level
implementations in the same host language and cover external systems, \eg, embedding the JVM-based
Esper~\cite{esper} in OCaml.  Embedding such systems additionally requires language bindings, \eg,
Ocaml-Java.\footnote{\url{www.ocamljava.org}} However, this is an orthogonal issue and we assume
that a suitable binding exists.  Once such a binding is in place, then \polyjoin{} can act as its
type-safe frontend for end users.

\myparagraph{Representing Context Dependency} In the embedding of \corrl{}
(\Cref{polyjoins:sec:type-safe-polyv}), we represented implicit context dependency by function
abstraction.
\citet{Bracevac:2018aa} propose an encoding of context dependency by means
of ambient, implicit parameters in the style of \cite{Lewis:2000aa}. These can be straightforwardly
encoded in a language with algebraic effects and handlers. However, this would require
compiler extensions, which makes this idea less portable.
In languages with compile-time ad-hoc polymorphism (\eg, \cite{Wadler:1989aa,Odersky:2018aa})
we could avoid the syntactic overhead of threading context through the join computation.

\myparagraph{On portability} In this paper, \polyjoin{} makes use of advanced OCaml features: GADTs,
phantom types, (first-class) modules, second-class higher-kinded polymorphism and Hindley-Milner
type inference.  However, it is portable to languages outside the ML family, as long as the target
host language can express some form of bounded polymorphism and type constructor polymorphism.
GADTs and phantom types can be encoded with interface/class inheritance, generics and subtyping. For
illustration, \Cref{sec:core-polyjoin-scala} shows a Scala version of \polyjoin{}.  We anticipate that programmers
can more conveniently and directly implement our ideas in languages with ad-hoc polymorphism, although
our Scala example does \emph{not} require it and works with local type inference~\cite{Pierce:1998aa}.  The more
``functional'' the host language, the easier it is to port and the more ``natural'' the DSL syntax
appears. We expect that \polyjoin{} should be expressible reasonably well even in modern Java
versions with lambdas. However, as other uses of tagless final/object algebras in Java
show~\cite{Biboudis:2015aa}, higher-kinded polymorphism has to be encoded indirectly, using the
technique by \citet{Yallop:2014aa}, which increases the amount of required boilerplate.

\myparagraph{Alternative Binder Representations} We investigated variants of \polyjoin{} that
support an $n$-ary
curried pattern notation, which has less syntactic noise than destructuring nested tuples:
\begin{lstlisting}[aboveskip=1pt,belowskip=1pt,numbers=none]
join ((from a) @. (from b) @. (from c) @. cnil)
      (fun x y z () -> (yield y))
\end{lstlisting}
We show the full tagless final specification for currying in \Cref{sec:curried-n-way}.
However, we found that the programming style is less convenient than the uncurried version, because
in the encoding that we considered, join has the following signature
\begin{lstlisting}[numbers=none,aboveskip=1pt,belowskip=1pt]
val join: ('shape, 'ps * 'res) ctx -> 'ps -> 'res shape repr
\end{lstlisting}
where the type parameter \lstinline|'ps| for the pattern body hides the fact that it is a
curried $n$-way function. To recover this information, structural recursion over the
context formation rules is required (as in \Cref{polyjoins:fig:monadic-cartesian}).
In contrast, we found that the uncurried style enabled a more natural, sequential
programming style, which works well with heterogeneous lists, as exemplified by the \corrl{}
case study (\Cref{polyjoins:sec:case-study:-cart}).

\subsection{Future Work}\label{polyjoins:sec:future-work}

\myparagraph{Disjunctions}
\polyjoin{} supports join patterns that are conjunctive, \ie, working on $n$-way products.
However, we did not address \emph{disjunctive} joins, yet, which are general notions of
patterns with cases (coproducts). For example, the CML choose combinator~\cite{Paykin:2016aa}
$$\lozenge A\otimes\lozenge B\Rightarrow\lozenge (A\otimes\lozenge B \oplus \lozenge A\otimes B)$$
is a linear logic version of pattern matching events with cases.

Another example for disjunction is
the Join Calculus~\cite{Fournet:1996aa}.  In future work, we would like to investigate a syntax
design that reasonably integrates disjunction with the event patterns of
\polyjoin{}. \citet{Rhiger:2009aa} shows that pattern cases are in principle expressible in terms of
typed combinators in higher-order functional languages.

\myparagraph{Shape Heterogeneity} The joins we cover in \polyjoin{} assume a fixed type
constructor/shape $S[\cdot]$,
but it seems useful to have a declarative join syntax that supports joining over heterogeneous
shapes $S_1[\cdot]$, \ldots, $S_n[\cdot]$ into an output shape $S_{n+1}[\cdot]$. \Eg,
$S_i \in\Set{\Type{Channel}, \Type{DB}, \Type{Future}, \Type{List},\ldots}$.
It seems promising to reduce the shape-heterogeneous case to the shape-homogeneous case,
by defining a common \lstinline|'a adapter| shape that encapsulates how to extract values
from heterogeneous shapes.

%%% Local Variables:
%%% mode: latex
%%% TeX-master: "report"
%%% End:

\section{Related Work}\label{polyjoins:sec:related-work}

\subsection{Language-Integrated Queries and Comprehensions}
Similarly to this work, Facebook's Haxl system \cite{Marlow:2014aa,Marlow:2016aa} recognizes that monad
comprehensions inhibit concurrency.  Their primary concern is reducing latency, exploiting
opportunities for data parallelism and caching in monadic queries that retrieve remote data
dependencies, without any observable change in the result. Thus, they analyze monad comprehensions
and automatically rewrite occurrences of monadic bind to applicative bind~\cite{McBride:2008aa},
whenever data parallelism is possible.  In contrast, our work is concerned with discerning and
correlating the arrival order of concurrent event notifications, which may lead to different
results.

In their T-LINQ/P-LINQ line of work, \citet{Cheney:2013aa} argue for having a quotation-based query
term representation having higher-order features, such as functional abstraction. Their system
guarantees that well-typed query terms always successfully translate to a SQL query and
normalization can avoid query avalanche.  \citet{Suzuki:2016aa} show that the T-LINQ approach by
Cheney et al.\ is definable in tagless final style, improving upon T-LINQ by supporting extensible
and modular definitions. \citet{Najd:2016aa} generalize the LINQ work by Cheney et al.\ to arbitrary
domain specific languages, using quotation, abstraction and normalization for reusing the host
language's type system for DSL types. They rely on Gentzen's subformula property to give guarantees
on the properties of the normalized terms by construction, \eg, absence of higher-order constructs
or loop fusion.  However, none of these works address general polyvariadic syntax forms as in
\polyjoin{}, outside of nested monadic bind. We consider it important future work to integrate
these lines of research with ours.

The join interface we propose in \Cref{polyjoins:def:joins} is an $n$-ary
version of Joinads, which underlie F\# computation
expressions~\cite{Petricek:2014ab,Syme:2011aa,Petricek:2011aa}.  The latter are a generalization of
monad comprehensions allowing for parallel binders, similar to \polyjoin{}. However in contrast to
our work, F\# computation expressions are not portable, requiring deep integration with the compiler
and are not as extensible and customizable.  \Eg, we support tight control of the pattern variable
signature, via shape translation and support new syntax forms. Furthermore, we support pattern
syntax designs for systems with implicit metadata.

\subsection{Polyvariadicity}

\citet{Danvy:1998aa} was the first to study polyvariadic functions to define a type-safe printf
function in pure ML and inspired many follow-up works, \eg, \cite{Fridlender:2000aa,Asai:2009aa}.
\citet{Rhiger:2009aa} would later define type-safe pattern matching combinators in
Haskell. Similarly to our work, Rhiger tracks the shape of pattern variable contexts at the
type-level and supports different binding semantics.

However, the above works rely on curried polyvariadic functions, while we choose an uncurried
variant, in order get ahold of the entire variable context in the tagless final join syntax. This
makes it easier to synthesize concurrent event correlation computations, because usually, all the
communicating components must be known in advance (cf. our discussion on alternative binder
representations in \Cref{polyjoins:sec:discussion}).  We postulate that usually, some variant of the
focusing problem (\Cref{polyjoins:sec:cart-prod-revis}) manifests in the concurrency case, where
each communication endpoint contributes one piece of information to a compound structure (\eg, the
cartesian product of mailboxes in Cartesius), but the code representing the contributions (\eg,
callback on the $i$-th event source) is position-dependent and heterogeneously-typed.

To the best of our knowledge, this work's combination of (1) tagless final, (2) abstract syntax with
polyvariadic signature, (3) GADT-based context formation and (4) type-level tracking of the variable
context shape is novel.  Importantly, the combination achieves modular polyvariadic definitions,
separating polyvariadic interfaces and polyvariadic implementations. Interactions with modules and
signatures as in \polyjoin{} have not been studied in earlier works on polyvariadicity. Some works
even point out the lack of modularity in their polyvariadic constructions, \eg, the numerals
encoding by \citet{Fridlender:2000aa}.  GADTs, which none of the previous works on polyvariadicity
consider, are the missing piece to close this gap, as they enable inversion/inspection of the
context formation.

\citet{Lindley:2008aa} encodes heterogeneously-typed many-hole contexts and context plugging for
type-safe embeddings of XML code in ML. His solution can express constraints on what kind of values
may be plugged into specific positions in a heterogeneous context. In our work, such constraints
would be useful for controlling that only certain combinations of restriction handlers
(\Cref{polyjoins:sec:extens-writ-restr}) can composed and applied. However, his representation of
heterogeneous sequences differs from ours and it remains open how to reconcile the two designs with
each other, which we consider interesting future work.

\citet{Weirich:2010aa,Weirich:2010ab} study forms of polyvariadicity
in the dependently-typed language Agda~\cite{Norell:2007aa} in terms of typed functions dependent on
peano number values. Dependent types grant more design flexibility for polyvariadic definitions, at the expense
of not being supported by mainstream languages. To the best of our knowledge, works on dependently-typed
polyvariadicity have not investigated modularity concerns so far.
\citet{McBride:2002aa} shows how to emulate dependently-typed definitions, including polyvariadic definitions,
in Haskell with multi-parameter type classes and functional dependencies. While more the above
approaches are more powerful, they are less portable and not as lightweight as \polyjoin{}.

\subsection{Higher-Order Abstract Syntax}

The idea of higher-order abstract syntax (HOAS) was first introduced by \citet{Huet1978}.  Later,
the paper by \citet{Pfenning:1988aa} popularized HOAS, which they implemented for the Ergo project,
a program design environment. HOAS eliminates the complications of explicit modeling of binders and
substitution in abstract syntax. One of their motivation was to simplify the formulation and ensure
correctness of syntactic rewriting rules in formal language development.  Similarly to this work,
they too recognize limitations of a purely curried specification style for $n$-ary
bindings and propose products for uncurried HOAS bindings with nested binary pairs, which required
changes to their implementation and unification algorithm. Our approach works ``out of the box''
with modern higher-order languages, both with HM and local type inference.  However, while their
motivation for uncurried binders were matters of formal syntax representation and manipulation, we
are motivated by denotation of syntax, \ie, a uncurried $n$-ary
binding forms satisfactorily enable concurrency implementations.

%%% Local Variables:
%%% mode: latex
%%% TeX-master: "report"
%%% End:

\section{Conclusion}\label{polyjoins:sec:conclusion}

We showed how to define type-safe language-integrated event patterns in terms of the tagless final
approach and explicit type-level representation of heterogeneous variable context in uncurried
style.  Our approach yields a practical and portable method to define extensible comprehension
syntax, without requiring dedicated compiler support.  In general, event patterns require a
non-monadic binding semantics and are thus not well-supported by state of the art
language-integrated query approaches.  We support patterns with an arbitrary number of event sources
and heterogeneous event types, while not requiring dependent types.  Event patterns and joins in
general, after all, are values which are dependent on the variable context.  Our encoding with
heterogeneous lists realizes this view in a direct way.

For event correlation, we found that computations follow a common pattern.  That is, evaluation must
focus ``in the middle'' of a heterogeneous shape and then collapse this shape
to express the correlation.  Equivalently, for each variable in the context, we must synthesize a
callback that relates the event notifications with the binder representation of the rest of the
context.

%%% Local Variables:
%%% mode: latex
%%% TeX-master: "report"
%%% End:

\begin{acks}
We thank Nada Amin, Oleg Kiselyov, Sam Lindley, and Jeremy Yallop for feedback and discussions on this work.
\end{acks}

\bibliography{report}
\clearpage
\appendix
\section{Optional, Supplementary Material}\label{sec:suppl-mater}

% \subsection{Polyvariadic Programming}\label{polyjoins:appendix:sec:polyv-progr}

% \input{fig-hlistfunctions}

\todo{cut}

\subsection{Extended Polyjoin}\label{polyjoins:appendix:sec:extended-polyjoin}

\begin{figure}[h]
%\centering
\begin{adjustbox}{valign=T,scale=.9}
\begin{lstlisting}[columns=fixed,aboveskip=0pt,belowskip=0pt]
module type ?Symantics? = sig
  type 'a shape (* $\commentcolor\Type{Shape}[\cdot]$ Constructor *)
  type meta     (* Event metadata type *)
  (* Judgments $\textit{\commentcolor (cf.~\Cref{polyjoins:fig:polyjoin-cartesius})}$: *)
  type 'a repr     (* $\commentcolor\hastype{\cdot}{A}$ *)
  type ('a,'b) pat (* $\commentcolor\formspatm{\vec{A}}{\cdot}{B}$ *)
  type ('a,'b) ctx (* combination of $\commentcolor\formsctx{\cdot}{A}\ \textit{and}\ \pvars{A}{B}$ *)
  type 'a var      (* $\commentcolor\isvar{\cdot}{A}$ *)
  type 'a ext      (* $\commentcolor\formsext{\vec{A}}{\cdot}$ *)
  (* Context Formation and Shape Translation: *)
  val from: 'a shape repr -> 'a var
  val cnil: (unit,unit) ctx
  val (@.): 'a var -> ('c, 'd) ctx -> ('a * 'c, 'a repr * 'd) ctx
  (* Contextual Extensions: *)
  val enil: unit -> 'a ext
  val (|++|): 'a ext -> 'a ext -> 'a ext
  (* Expressions and Patterns: *)
  val mmerge: meta repr -> meta repr -> meta repr
  val yield: 'a repr -> ('c,'a) pat
  val where: bool repr -> ('c,'a) pat -> ('c,'a) pat
  val join: ('a, 'b) ctx -> 'a ext -> ('b -> ('a,'c)) pat) -> 'c shape repr
end
\end{lstlisting}
\end{adjustbox}
\myhack{\vspace{-20pt}}
\caption{Tagless Final Representation of Extended \polyjoin{}.}\label{polyjoins:fig:extendedsymantics}
\end{figure}

\subsection{Core PolyJoin in Scala}\label{sec:core-polyjoin-scala}
\begin{lstlisting}[language=Scala]
import scala.language.higherKinds
//Module signatures become traits/interfaces
trait Symantics {
    type Ctx[A]
    type Var[A]
    type Repr[A]
    type Shape[A]
    def from[A](shape: Repr[Shape[A]]): Var[Repr[A]]
    val cnil: Ctx[Unit]
    def ccons[A,B](v: Var[A], ctx: Ctx[B]): Ctx[(A,B)]
    def join[A,B](ctx: Ctx[A])(pattern: A => Repr[Shape[B]]): Repr[Shape[B]]
    def yld[A](x: Repr[A]): Repr[Shape[A]]
    def pair[A,B](fst: Repr[A], snd: Repr[B]): Repr[(A,B)]
  }

//For testing, extend the language with a syntax to lift lists of values to shape representations
trait SymanticsPlus extends Symantics {
   //note: A* means "variable number of A arguments"
   def lift[A](xs: A*): Repr[Shape[A]]
}

//Expression functors become functions/methods with path-dependent types
def test(s: SymanticsPlus) //infers path-dependent type s.Repr[s.Shape[(Double, (Int, String))]]
= {
  import s._
  //Note: with more effort, the syntax for constructing contexts can be made prettier, in infix notation
  val ctx = ccons(from(lift(1,2,3)), ccons(from(lift("one", "two")), ccons(from(lift(3.0,2.0,1.0)), cnil)))
  //Notational convenience is similar to the OCaml version
  join (ctx) { case (x,(y,(z,()))) =>
     yld(pair(z,pair(x,y)))
  }
  /* this wouldn't type check:
   join (ctx) { case (x,(y,())) =>
     yld(pair(z,pair(x,y)))
   }
   */
}

//Example: sequential cartesian product over lists
object ListSymantics extends SymanticsPlus {
  //GADT becomes class hierarchy
  sealed abstract class Ctx[A]
  //GADT constructors become case classes/objects
  case object CNil extends Ctx[Unit]
  case class CCons[A,B](hd: Var[A], tl: Ctx[B]) extends Ctx[(A,B)] //refinement controlled by extends clause
  sealed abstract class Var[A]
  case class Bind[A](shape: Repr[Shape[A]]) extends Var[Repr[A]]
  override type Repr[A] = A
  override type Shape[A] = List[A]
  override def from[A](shape: Repr[Shape[A]]) = Bind(shape)
  override val cnil = CNil
  override def ccons[A, B](v: Var[A], ctx: Ctx[B]) = CCons(v,ctx)
  override def lift[A](xs: A*) = List(xs:_*)
  override def yld[A](x: Repr[A]) = List(x)
  override def pair[A, B](fst: Repr[A], snd: Repr[B]) = (fst,snd)

  /* This join implements the monadic sequential semantics */
  override def join[A, B](ctx: Ctx[A])(pattern: A => Repr[Shape[B]]) = ctx match {
    case CNil => pattern ()
    case CCons(Bind(xs), tl) =>
      xs.flatMap { x =>
        join (tl) { tuple => pattern (x,tuple) }
      }
  }
}

test(ListSymantics)
/* prints res0: ListSymantics.Repr[ListSymantics.Shape[(Double, (Int, String))]] =
List((3.0,(1,one)), (2.0,(1,one)), (1.0,(1,one)), (3.0,(1,two)), (2.0,(1,two)),
(1.0,(1,two)), (3.0,(2,one)), (2.0,(2,one)), (1.0,(2,one)), (3.0,(2,two)),
(2.0,(2,two)), (1.0,(2,two)), (3.0,(3,one)), (2.0,(3,one)), (1.0,(3,one)),
(3.0,(3,two)), (2.0,(3,two)), (1.0,(3,two))) */
\end{lstlisting}

\subsection{Curried $n$-way Joins}\label{sec:curried-n-way}

\begin{lstlisting}
module type ?Curried? = sig
  type 'a repr
  type 'a shape
  type 'a pat
  type ('a,'b) ctx
  type ('a,'b) var
  val from: 'a shape repr -> ('a, 'a repr) var
  val cnil: (unit, (unit -> 'a pat) * 'a) ctx
  val (@.): ('a,'b) var -> ('c, 'd * 'e) ctx -> ('a * 'c, ('b -> 'd) * 'e) ctx
  val yield: 'a repr -> 'a pat

  (* Problem: it is not apparent that 'ps is a pattern abstraction, which forces
     a structurally inductive programming style over the context formation.  *)
  val join: ('shape, 'ps * 'res) ctx -> 'ps -> 'res shape repr
end

module ?TestCurried?(?C?: ?Curried?) = struct
  open ?C?
  let test a b c =
    join ((from a) @. (from b) @. (from c) @. cnil)
      (fun x y z () -> (yield y))
end
\end{lstlisting}

%\subsection{Combine Latest Semantics}
%\label{polyjoins:appendix:sec:comb-latest-semant}
%\input{fig-signals-ex}

%%% Local Variables:
%%% mode: latex
%%% TeX-master: "report"
%%% End:

\end{document}